\documentclass[aps, prd, nofootinbib, preprintnumbers,showkeys,preprint]{revtex4}
\usepackage{amsmath}
\usepackage{amssymb}
\allowdisplaybreaks

\usepackage[thicklines]{cancel}
\usepackage{makeidx}
\usepackage{amsfonts}
\usepackage[usenames,dvipsnames]{pstricks}
\usepackage{subfigure}
\usepackage{epsfig}
\usepackage{pst-grad} 
\usepackage{mathtools}
\usepackage[colorlinks,hyperindex]{hyperref}
\usepackage{array}
\usepackage{tikz}
\usepackage{float}
\usetikzlibrary{decorations.markings,math,shapes.misc,patterns,arrows.meta,calc,patterns,angles,quotes,intersections}
\usepackage{allpurpose}
\newcommand {\nn}{\nonumber}

\hypersetup
{colorlinks,%
citecolor=black,%
linkcolor=black,%
urlcolor=black,%
}

\begin{document}

\title{Deflection in higher dimensional spacetime and asymptotically non-flat spacetimes}

\author{Jinhong He}
\thanks{These authors contributed equally to this work. }
\affiliation{School of Physics and Technology, Wuhan University, Wuhan,  430072, China}

\author{Qianchuan Wang}
\thanks{These authors contributed equally to this work. }
\affiliation{School of Physics and Technology, Wuhan University, Wuhan,  430072, China}

\author{Qiyue Hu}
\affiliation{School of Physics and Technology, Wuhan University, Wuhan,  430072, China}

\author{Li Feng}
\affiliation{School of Physics and Technology, Wuhan University, Wuhan,  430072, China}

\author{Junji Jia}
\email[Corresponding author:~]{junjijia@whu.edu.cn}
\affiliation{Center for Astrophysics $\&$ MOE Key Laboratory of Artificial Micro- and Nano-structures, School of Physics and Technology, Wuhan University, Wuhan, 430072, China}

\begin{abstract}
Using a perturbative technique, in this work we study the deflection of null and timelike signals in the extended Einstein-Maxwell spacetime, the Born-Infeld gravity and the charged Ellis-Bronnikov (CEB) spacetime in the weak field limit. The deflection angles are found to take a (quasi-)series form of the impact parameter, and automatically takes into account the finite distance effect of the source and observer. The method is also applied to find the deflections in CEB spacetime with arbitrary dimension. It's shown that to the leading non-trivial order, the deflection in some $n$-dimensional spacetimes is of the order $\mathcal{O}(M/b)^{n-3}$. We then extended the method to  spacetimes that are asymptotically non-flat and studied the deflection in a nonlinear electrodynamical scalar theory. The deflection angle in such asymptotically non-flat spacetimes at the trivial order is found to be not $\pi$ anymore. In all these cases, the perturbative deflection angles are shown to agree with numerical results extremely well. The effects of some nontrivial spacetime parameters as well as the signal velocity on the deflection angles are analyzed.
\end{abstract}

\keywords{deflection angle, perturbative method, timelike signal, high dimensional spacetime}
\date{\today}

\maketitle

\section{Introduction}

The deflection and gravitational lensing (GL) of light rays are important tools in astronomy. The former historically contributed significantly to the acceptance of general relativity (GR) by scientists  \cite{Dyson:1920cwa}. And the latter is being used to find exoplanets \cite{Mao:1991nt,SupernovaCosmologyProject:1993faz}, to measure the mass distribution of galaxies and their clusters \cite{Hoekstra:2000ux,Gray:2001zx}, and to test theories beyond GR \cite{Hoekstra:2008db,Joyce:2016vqv}.
In recent years, with the observation of gravitational wave (GW) \cite{LIGOScientific:2016aoc} and the black hole shadow \cite{EventHorizonTelescope:2019dse,EventHorizonTelescope:2019ggy,EventHorizonTelescope:2022xnr}, the deflection and GL of GW, as well as the deflection and GL of light rays in the strong field limit, have drawn enormous amount of attention.

On the other hand, after the discovery of cosmic rays \cite{Hess:1912srp} and the SN 1987A neutrinos \cite{Kamiokande-II:1987idp,Bionta:1987qt}, especially after the confirmation of the extrasolar origin of the former
\cite{Baade:1934zex,Baade:1934zex2}
and the nonzero mass of the latter \cite{Cleveland:1998nv,Super-Kamiokande:1998kpq}, people become aware that massive signals such as cosmic rays and neutrinos from various sources can also experience gravitational deflection and be messengers of GL. One particularly encouraging progress in this direction is the discovery of GLed supernovas in recent years \cite{Kelly:2014mwa,Goobar:2016uuf}.

To theoretically investigate the deflection of these signals, recently two methods have been intensively used. One is to use the Gauss-Bonnet theorem method \cite{Glavan:2019inb,Gibbons:2008rj,Werner:2012rc}, which has been developed to handle both null and timelike signals
\cite{Werner:2012rc,Ono:2017pie,Li:2019qyb,Li:2021xhy}, and to take into account the finite distance effect of the source and detector \cite{Ishihara:2016vdc,Ishihara:2016sfv,Li:2020ozr}, as well as the electromagnetic force \cite{Crisnejo:2019xtp,Li:2020ozr,Li:2021xhy}. The other method is the perturbative method developed by some authors of this paper. The method can also calculate the deflection of both null and timelike signals \cite{Jia:2020xbc} and has been extended to include the finite distance effect \cite{Huang:2020trl} and the extra kind of force \cite{Xu:2021rld,Zhou:2022dze} too. Moreover, the perturbative method can be used in arbitrary stationary and axisymmetric spacetimes \cite{Jia:2020xbc,Huang:2020trl}, as well as in the strong field limit \cite{Jia:2020qzt}.

There are two folds of motivations of this work. The first is to apply the perturbative method previously developed to other interesting spacetimes, namely the extended Einstein-Maxwell spacetime \cite{Cano:2020ezi}, the Born-Infeld  gravity \cite{Jana:2015cha} and the charged Ellis-Bronnikov (CEB) spacetime \cite{Nozawa:2020wet}, to study the effect of the various spacetime parameters on the deflection of signals in these spacetimes. The second motivation is to test whether the perturbative method can be extended to treat rare spacetimes, particularly the higher dimensional ones and asymptotically non-flat ones.
As we will see, it turns out the extension is actually quite simple and we will apply it to the higher dimensional CEB spacetime and a nonlinear electrodynamical scalar (NES) theory.

The paper is organized as follows. In Sec. \ref{sec:method} we extend the perturbative method that was previously developed to arbitrarily high dimensional spacetime. In Sec. \ref{sec:applications} we apply the method to three four-dimensional spacetimes that are asymptotically flat, and study the effects of the spacetime parameters on the deflection of both null and timelike signals. We emphasize that to our knowledge, the deflections in these spacetimes in the weak field limit were not studied before. Moreover, in Subsec. \ref{subsec:3}, we also use the perturbative method to study the deflection in the higher dimensional CEB spacetime. In Sec. \ref{sec:asynonm}, the perturbative method is extended to the asymptotically non-flat case and used to study the NES theory. We also point out a prominent feature of the deflection angle in this kind of spacetimes. We conclude the work with a short discussion in Sec. \ref{sec:conc}.
Throughout this work, we adopt the natural unit system $G=c=1$ and the most plus metric convention.

\section{The perturbative method\label{sec:method}}

The perturbative method to calculate the deflection angle in the static and spherically symmetric spacetimes in the weak field limit was initiated in Ref. \cite{Jia:2020xbc} and further formalized in Ref. \cite{Huang:2020trl}. There, the metric functions were assumed to allow asymptotic expansions into integer power series of the radius. In this section, we will first extend the main procedure of this method to arbitrarily high dimensional spacetimes and then further extend it to asymptotically non-flat metrics in Sec. \ref{sec:asynonm}.

Static and spherically symmetric spacetimes in $n$-dimensional ($n\geq 4$) spacetimes can always be described by the line element
\begin{equation}
\dd s^2=-A(r) \dd t^2+B(r)\dd r^2+C(r) (\dd\theta^2+\sin^2\theta\dd\phi^2+\cos^2\theta \dd \Omega_{n-4}^2), \label{eq:sssmetric}
\end{equation}
where
\begin{align}
\dd\Omega_{n-4}^2=\dd\chi_1^2+\sin^2\chi_1\dd\chi_2^2+\cdots+\prod_{i=1}^{n-5}\sin^2\chi_i\dd\chi_{n-4}^2
\end{align}
and $(t,~r,~\theta,~\phi,~\chi_1,~\chi_2,~\cdots,~\chi_{n-4})$ are the coordinates and $A(r),~B(r)$ and $C(r)$ are the metric functions of $r$ only. The asymptotic flatness of the spacetime often allows the following asymptotic expansion of the metric functions
\be
A(r) =1+\sum_{n=1} \frac{a_n}{r^n},\ B(r) =1+\sum_{n=1} \frac{b_n}{r^n},\ \frac{C(r)}{r^2}=1+\sum_{n=1} \frac{c_n}{r^n}, \label{eq:abcform}
\ee
where $a_n$, $b_n$ and $c_n$ are finite constants. Although locally we can always set $C(r)=r^2$, there are occasions that $C(r)$ is transformed to other forms and therefore we will keep its general form as in \eqref{eq:abcform} for now. Due to spherically symmetry of the spacetime, we need only to consider the particle trajectory on the equatorial plane  ($\theta=\pi/2$ and $\chi_i$=constant). The geodesic equations associated with a test particle in this plane then can be readily obtained as
\begin{subequations}
\label{eq:gdeq}
\begin{align}
\dot{t}&=\frac{E}{A(r)},\label{eq:gdeq1}\\
\dot{\phi}&=\frac{L}{C(r)},\label{eq:gdeq2}\\
\dot{r}^2&=\frac{1}{B(r)}\left(\kappa-\frac{E^2}{A(r)}+\frac{L^2}{C(r)}\right),\label{eq:gdeq3}
\end{align}
\end{subequations}
where the dot stands for the derivative with respect to the affine parameter. $L$ and $E$ are respectively the conserved angular momentum and energy per unit mass of the signal, and $\kappa = 0$ and $1$ respectively for null and timelike signals.

\begin{figure}[htp!]
\centering
\includegraphics[width=0.45\textwidth]{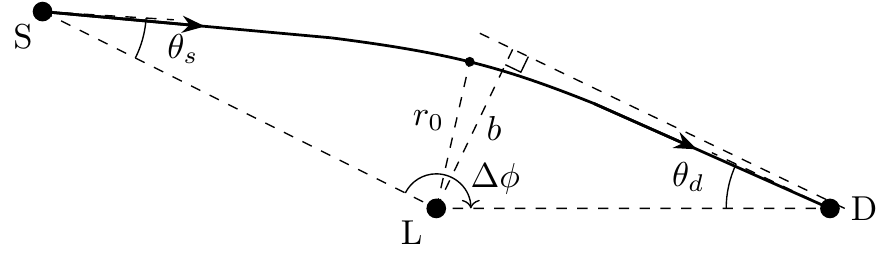}
\caption{The trajectory from the source S at radius $r_s$ to the detector D at $r_d$, passing by the lens L. The impact parameter and closest approach are marked as $b$ and $r_0$ respectively. \label{fig:defgl}}
\end{figure}

From these equations, one can express the deflection angle $\Delta \phi$ from the source at coordinate $(r_s,\phi_s)$ to the detector at $(r_d,\phi_d)$ (see Fig. \ref{fig:defgl}) as \cite{Huang:2020trl,Liu:2020wcu}
\be
\Delta \phi=\lsb \int_{r_0}^{r_s}+\int_{r_0}^{r_d}\rsb \sqrt{\frac{B}{C}}\frac{L}{\sqrt{(E^2/A-\kappa )C-L^2}}\dd r,
\label{eq:deltaphi}
\ee
where $r_0$ is the closest approach of the trajectory and can be solved from the condition $\dot{r}|_{r=r_0}=0$. Using Eq. \eqref{eq:gdeq3}, this condition can also establish a relation between $L$ and $r_0$
\be
L=\sqrt{[E^2-\kappa A(r_0)]C(r_0)/A(r_0)}. \label{eq:linr0}
\ee

In an asymptotically flat spacetime, $L$ and $E$ can also be expressed as
\be
L=|\vecr\times \vp|=\frac{v}{\sqrt{1-v^2}}b,~E=\frac{1}{\sqrt{1-v^2}}, \label{eq:leandbv}
\ee
where $v$ is the signal velocity at infinity and $b$ is the impact parameter.
Solving $b$ from Eq. \eqref{eq:leandbv} and using Eq. \eqref{eq:linr0}, we have
\bea
\frac{1}{b}&=&\frac{\sqrt{E^2-\kappa}}{\sqrt{E^2-\kappa A(r_0)}}\sqrt{\frac{A(r_0)}{C(r_0)}} \label{eq:pfuncdefp}\\
&\equiv& p\lb \frac{1}{r_0}\rb \label{eq:pfuncdef}
\eea
where in the last step the right-hand side of Eq. \eqref{eq:pfuncdefp} is defined as a function $p$ of $1/r_0$. For later purpose, we will denote the inverse function of $p(x)$ as $q(x)$. We point out that as long as the metric functions are known, both functions $p(x)$ and $q(x)$ can be known too (at least perturbatively).

To calculate $\Delta \phi$ in \eqref{eq:deltaphi}, in Ref. \cite{Huang:2020trl,Liu:2020wcu} we proposed the change of variable from $r$ to $u$ using the relation
\be
r=1/q\lb \frac{u}{b}\rb . \label{eq:rtoucov}
\ee
Substituting Eq. \eqref{eq:rtoucov} into Eq. \eqref{eq:deltaphi}, it is not too difficult to verify that it is transformed to
\be
\Delta\phi=\lsb \int_{\sin\theta_s}^{1}+\int_{\sin\theta_d}^{1}\rsb y\lb \frac{u}{b} \rb \frac{\dd u}{\sqrt{1-u^2}}, \label{eq:phiinu}
\ee
where
\be
y\lb \frac{u}{b}\rb = \sqrt{\frac{B(1/q)}{C(1/q)}}\frac{1}{p'(q)q^2}\frac{u}{b}, \label{eq:ydef}
\ee
and
\be
\theta_s=\arcsin\lsb b\cdot p\lb \frac{1}{r_s}\rb \rsb,~\theta_d=\arcsin\lsb b\cdot p\lb \frac{1}{r_d}\rb \rsb\label{eq:thetasddef}
\ee
are actually the apparent angles of the signal at the source and detector respectively.

In the weak field limit, $ y\lb u/b\rb $ in Eq. \eqref{eq:phiinu} can be expanded into a power series of $u/b$, i.e.,
\be
y\lb \frac{u}{b}\rb =\sum_{n=0}^\infty y_n \lb  \frac{u}{b}\rb^n, \label{eq:yexp}
\ee
where $y_n$ can be expressed in terms of coefficients in the asymptotic expansions \eqref{eq:abcform} of the metric. The first few of them are
\begin{subequations}
\label{eq:yncoeffall}
\begin{align}
y_0=&1,\\
y_1=&\frac{b_1}{2}-\frac{a_1}{2 v^2},\\
y_2=&\frac{4(c_2+b_2)-(c_1-b_1)^2 }{8}+\frac{a_1(2a_1-c_1-b_1)-2a_2}{2v^2}.
\end{align}
\end{subequations}
Higher order coefficients are also easily obtained and can be seen in Appendices of Ref. \cite{Huang:2020trl,Liu:2020wcu}.
From Eq. \eqref{eq:yncoeffall}, it is seen that for the order $n$ coefficient $y_n$, only the metric expansion coefficients up to order $n$, i.e. $a_n,~b_n$ and $c_n$, contribute.

Substituting Eq. \eqref{eq:yexp} into Eq. \eqref{eq:phiinu}, $\Delta\phi$ becomes a sum containing a series of integrals of the form
\be
I_n(\theta_s,\theta_d)=\lsb \int_{\sin\theta_s}^{1}+\int_{\sin\theta_d}^{1}\rsb \frac{u^n}{\sqrt{1-u^2}}\dd u,~ (n=0,1,\cdots). \label{eq:indef}
\ee
Integrals of this type can always be carried out and their results are present in Eq. \eqref{eq:inres2} in Appendix \ref{sec:appd}. Therefore, the above change of variables and series expansion guarantee that we can obtain the following quasi-inverse power series form of the deflection angle
\be
\Delta\phi =\sum_{n=0}^\infty y_n\frac{I_n(\theta_s,\theta_d)}{b^n} . \label{eq:dphifinal}
\ee
In the large $r_s$ and $r_d$ limit, from Eq. \eqref{eq:thetasddef} we see that $\theta_s,~\theta_d$ are small and then one can expand $I_n(\theta_s,\theta_d)$ as series of $b/r_s$ and $b/r_d$ using Eq. \eqref{eq:i00res}. After this, $\Delta \phi$ in Eq. \eqref{eq:dphifinal} becomes pure series of $(M/b)$ and $(b/r_{s,d})$. To the first few orders, it is
\begin{align}
\Delta\phi&=\sum_{i=s,d}\frac{\pi}{2}+\frac{1}{2b}\left(b_1-\frac{a_1}{v^2}\right)-\frac{\pi}{32b^2}\left\{\left(b_1-c_1\right)^2-4(b_2+c_2)+\frac{8}{v^2}\left[\frac{ a_1 \left(b_1+c_1\right)}{2}- a_1^2+ a_2\right]\right\}\nn\\
&+\frac{1}{b^3}\left\{ \frac{1}{24}\left[-2 b_1^2 c_1-4 b_1 \left(b_2-2 c_2\right)+8 \left(b_2 c_1+b_3+2 c_3\right)+b_1^3\right]\right.\nn\\
&+\frac{1}{8v^2}\left[4 a_1^2 \left(b_1+2 c_1\right)+a_1 \left(16 a_2-4 b_1 c_1-4 \left(b_2+2 c_2\right)+b_1^2\right)-4 a_2 \left(b_1+2 c_1\right)-8 a_1^3-8 a_3\right]\nn\\
&\left.+\frac{1}{8v^4}\left[a_1 \left(a_1 \left(b_1+2 c_1\right)-4 a_1^2+4 a_2\right)\right]+\frac{a_1^3}{24 v^6}\right\}\nn\\
&-\frac{b}{r_i}+\frac{b}{4r_i^2}\left(-b_1+2c_1-\frac{a_1}{v^2}\right)-\frac{b^3}{6r_i^3} +\mathcal{O}\left[\left(\frac{M}{b}\right)^4,  \left(\frac{b}{r_i}\right)^4\right]. \label{eq:dphiwithfd}
\end{align}
In the infinite $r_s,~r_d$ limit, this becomes
\begin{equation}
\Delta\phi=\pi+\left(b_1-\frac{a_1}{ v^2}\right)\frac{1}{b}+\left(\frac{4(b_2+c_2)-(b_1-c_1)^2 }{4}+\frac{a_1(2a_1-b_1-c_1)-2a_2}{v^2}\right)\frac{\pi}{4b^2}+\calco\left(\frac{1}{b}\right)^3.\label{eq:definf}
\end{equation}

We note particularly from Eq. \eqref{eq:definf} that, for the coefficient of the deflection of order $(M/b)^k~(k\geq 1)$, only metric coefficients up to order $a_k,~b_k$ and $c_k$ appear but not higher order ones.
For deflection in spacetimes with dimension $n\geq5$, we know that some of these spacetimes \cite{Myers:1986un,Belhaj:2020rdb,Singh:2023ops} (not all though, see \cite{Mazharimousavi:2008ap}
)
have non-trivial metric expansion coefficients starting only from order $n-3$, i.e., $a_1=\cdots=a_{n-4}=b_1=\cdots=b_{n-4}=c_1=\cdots=c_{n-4}=0$.
For such spacetimes, clearly the first non-trivial term of the deflection with infinite $r_s,~r_d$ will be of order $\mathcal{O}(M/b)^{n-3}$. If $n$ is much larger than four, this will force the deflection to be extremely small. If we compare this with the finite distance correction, we see from Eq. \eqref{eq:dphiwithfd} that the latter always starts from the order $\mathcal{O}(b/r_{s,d})^1$, which could be larger than the $\mathcal{O}(M/b)^{n-3}$ order term in true gravitational lensing. This reminds us the importance of the finite distance effect in higher dimensional spacetimes. We will see a detailed example in Sec. \ref{subsec:3} for this phenomenon.

\section{Applications to particular spacetimes\label{sec:applications}}

In this section, we will directly apply the above method to some known spacetimes to check the validity of Eq. \eqref{eq:dphifinal}, and more importantly, to reveal how any parameters of the spacetime and the particle velocity will affect the deflections.

\subsection{Extended Einstein-Maxwell Spacetime\label{subsec:1}}

The extended Einstein-Maxwell theory describes a charged spacetime without the central singularity \cite{Cano:2020ezi}. Its line element is given by Eq.  \eqref{eq:sssmetric} with the following metric functions
\begin{align}
A(r) =\frac{1}{B(r)}=\frac{r^4 (r^2-2Mr+Q^2)+\alpha  Q^2(3r^2+2\alpha)} {r^6+\alpha Q^2(r^2+2\alpha)},~C(r)=r^2. \label{eq:abcern}
\end{align}
Here $M$ and $Q$ are respectively the spacetime mass and charge, while $\alpha$ is a scale parameter with a length square dimension.
Expanding these metric functions asymptotically, we have up to the fourth order
\begin{subequations}
\begin{align}
&A(r)=1 - \frac{2 M}{r} + \frac{Q^2 }{r^2} + \frac{2 Q^2\alpha}{r^4 } +\mathcal{O}(r)^{-5}, \\
&B(r)=1 +\frac{2 M}{r} + \frac{4 M^2 - Q^2}{r^2} + \frac{8 M^3 - 4 M Q^2}{r^3}  + \frac{16 M^4 - 12 M^2 Q^2+ Q^4 - 2 Q^2 \alpha}{r^4}+\mathcal{O}(r)^{-5},\\
&\frac{C(r)}{r^2}=1. \label{eq:eemexp}
\end{align}
\end{subequations}
Reading off the coefficients $a_i,~b_i$ and $c_i$ and substituting into Eq. \eqref{eq:yncoeffall}, we get the coefficients $y_n$ up to the fourth order in the deflection angle $\Delta\phi_E$ in this spacetime, as
\begin{subequations}
\label{eq:ynern}
\begin{align}
y_{E,0}=&1,\\
y_{E,1}=&M\left(1+\frac{1}{v^2}\right),\\
y_{E,2}=&M^2\left(\frac32+\frac{6}{v^2}\right)-Q^2\left(\frac{1}{2}+\frac{1}{v^2}\right),\\
y_{E,3}=&M^3\left(\frac{5}{2}+\frac{45}{2v^2}+\frac{15}{2v^4}-\frac{1}{2v^6}\right)-M Q^2\left(\frac{3}{2}+\frac{9}{v^2}+\frac{3}{2v^4}\right),\\
y_{E,4}=&M^4\left(\frac{35}{8}+\frac{70}{v^2}+\frac{70}{v^4}\right)-M^2Q^2\left(\frac{15}{4}+\frac{45}{v^2}+\frac{30}{v^4}\right)+Q^4\left(\frac{3}{8}+\frac{3}{v^2}+\frac{1}{v^4}\right)\nonumber\\
&-Q^2\alpha\left(1+\frac{4}{v^2}\right). \label{eq:ynern4}
\end{align}
\end{subequations}
Substituting these into Eq. \eqref{eq:dphifinal}, $\Delta\phi_E$ is found to be
\be
\Delta\phi_E=\sum_{n=0}^\infty y_{E,n} \frac{I_n(\theta_s,\theta_d)}{b^n}, \label{eq:dphires1}
\ee
where $I_n(\theta_s,\theta_d)~(n=0,1,\cdots)$ are given in Eq. \eqref{eq:inres2}. Note that higher than the fourth order results for the coefficients are easily obtainable too but not shown here for their excessive length.

\begin{figure}[htp!]
\centering
\includegraphics[width=0.45\textwidth]{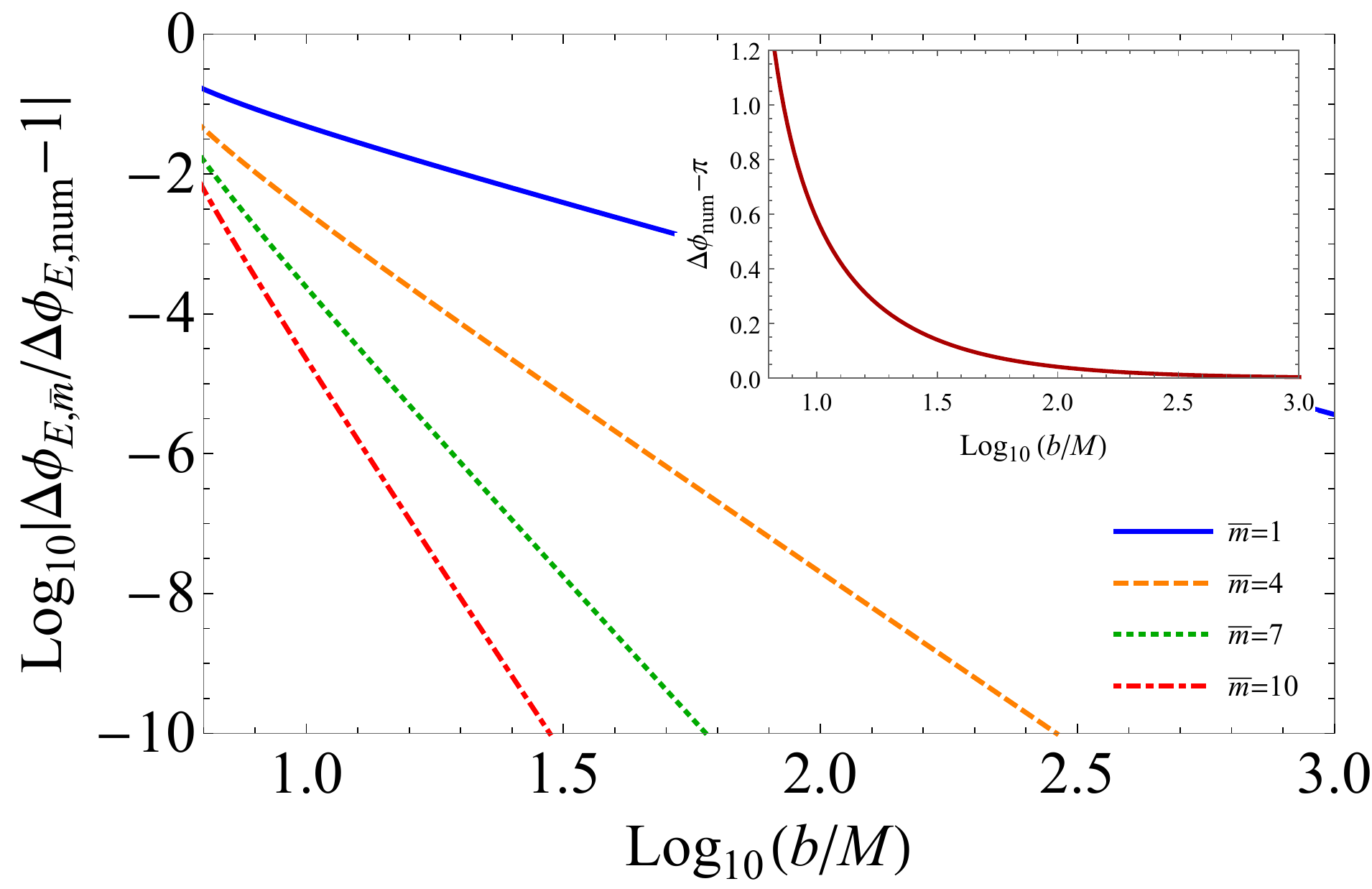}
\includegraphics[width=0.45\textwidth]{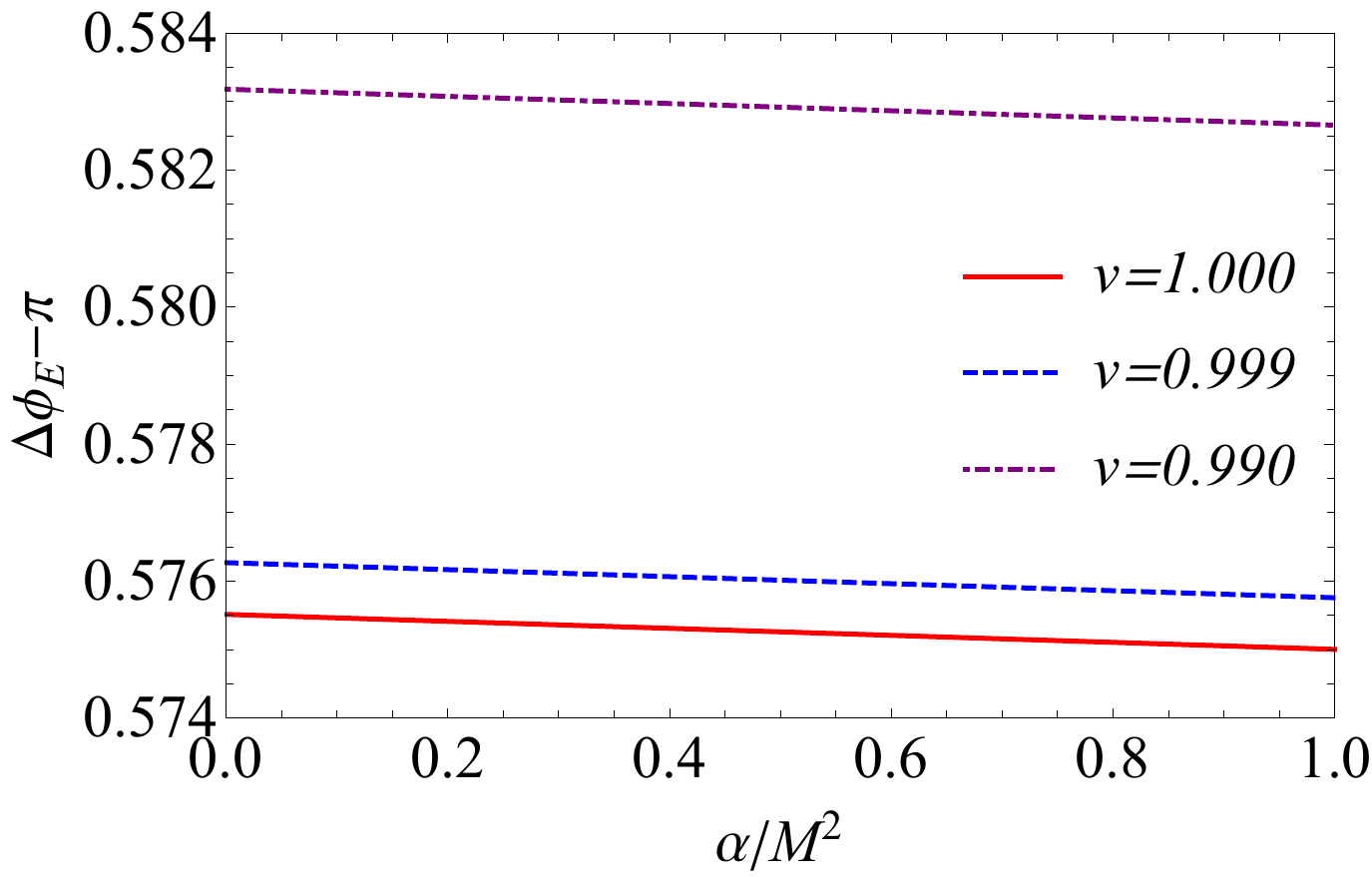}\\
(a)\hspace{7cm}(b)
\caption{(a) Difference between the perturbative $\Delta\phi_{E,\bar{m}}$ and the numerical result $\Delta\phi_{E,\mathrm{num}}$ as a function of $b/M$ from 7 to $10^3$. The inset shows the deflection angle itself for this range of $b$. Other parameters used are $Q=M/2,~v=1-10^{-2}, ~r_s=r_d=10^6M$ and $\alpha=10^{-2}M^2$. (b) $\Delta \phi_{E}-\pi$ as a function of $\alpha$ for $b=10M$ and three values of $v=1,~0.999,~0.990$. Other parameters are the same as in (a). \label{phi-b}}
\end{figure}

To check the correctness of $\Delta\phi_E$, we define a truncated $\Delta\phi_{E,\bar m}$ as the sum of up to the $\bar{m}$-th order
\be
\Delta\phi_{E,\bar{m}}=\sum_{n=0}^{\bar{m}} y_n \frac{I_n(\theta_s,\theta_d)}{b^n}\quad
(\bar{m}=1,~2,~\cdots),
\ee
and then in Fig. \ref{phi-b} (a) we plot the difference between $\Delta\phi_{E,\bar{m}}$ and the numerical integration result $\Delta\phi_{E,\mathrm{num}}$ of the original Eq. \eqref{eq:deltaphi} using the metric \eqref{eq:abcern}.
As long as the numerical integration is done with high enough precision, its result can be thought as the true deflection of the trajectory and all perturbative results can be compared against it. We then see from the figure that for any fixed impact parameter $b$, as the truncation order increases, the perturbative result approaches the true value exponentially. Moreover, we see that as $b$ increases, $\Delta \phi_E$ decreases (see the inset) as dictated by the quasi-inverse power series of the form \eqref{eq:dphires1}, and the accuracy for each $\Delta\phi_{E,\bar m}$ also increases. At about $b=10^3M$, the relative difference between $\Delta\phi_{E,\bar{4}}$ and $\Delta\phi_{E,\mathrm{num}}$ is less than $10^{-10}$. Note that the relative difference between $\Delta\phi_{E,\bar{10}}$ and $\Delta_{E,\mathrm{num}}$ at impact parameter as small as $b=10M$ is still quite small, while the $\Delta\phi_{E}$ itself at this point is already 0.6, not truly a weak deflection anymore. All the above features and comparison show that the  $\Delta\phi_{E}$ found in Eq. \eqref{eq:dphires1} can approximate the true deflection angle to very high precision as long as the truncation order is high enough.

With the correctness of Eq. \eqref{eq:dphires1} confirmed, we can now use it to study the effects of the scale parameter $\alpha$ and the kinetic variable $v$ on the deflection angle. Here we will concentrate on the effect of the parameter $\alpha$ but not charge $Q$ because the latter was well studied in Reissner-Nordstr\"{o}m (RN) spacetime \cite{Pang:2018jpm}.
Eq. \eqref{eq:ynern4} shows that $\alpha$ only appear from the fourth order $y_4$ in $\Delta\phi_E$ and therefore it is expected that in general, its effect to the deflection will be very small as long as its size does not exceed unity, i.e.,  $\alpha\lesssim \mathcal{O}(M^2)$. Moreover, it is also clear that the smaller the impact parameter $b$,  the larger the effect of $\alpha$ on the overall size of $\Delta\phi$.
In Fig. \ref{phi-b} (b) we plot the variation of $\Delta\phi_E$ as a function of $\alpha$ for $b=10M$. It is clear that as $\alpha$ increases, the deflection angle decreases monotonically. This is consistent with Eq. \eqref{eq:ynern4} in which the term involving $\alpha$ has a negative sign. It is also seen that the larger the velocity $v$, the smaller the deflection, which is also in accord with the intuition from Newtonian mechanics.

\subsection{The Born-Infeld gravity \label{subsec:2}}
The Born-Infeld gravity describes gravity coupled with Born-Infeld electrodynamics \cite{Jana:2015cha}. Its
metric functions for a special type are given by \cite{Jana:2015cha}
\begin{subequations}
\begin{align}
A(r)&=\lb 1-\frac{4\eta Q^2}{(r^2+\sqrt{r^4+4\eta Q^2})^2}\rb \lb 1-\frac{2\sqrt{2}M}{\sqrt{r^2+\sqrt{r^4+4\eta Q^2}}}+\frac{2Q^2}{r^2+\sqrt{r^4+4\eta Q^2}}\rb,\\
B(r)&=\frac{2r^2}{(r^2+\sqrt{r^4+4\eta Q^2})A(r)},\\
\frac{C(r)}{r^2}&=1. \label{eq:bimetrics}
\end{align}
\end{subequations}

Here $M$ and $Q$ are respectively the mass and charge of the spacetime and $\eta>0$ characterizes the relative scale of the electromagnetic and gravity sectors of the theory.
When $\eta=0$, this reduces to the classical RN spacetime. Therefore, $\eta$ can be regarded as a parameter that measures the deviation of the gravity from the RN spacetime.  Only when $|Q|\leq M$ and $0<\eta<\eta_+$ where $\eta_\pm=M^4\lb 1\pm \sqrt{1-Q^2/M^2}\rb^4/Q^2$ will this metric describe a BH spacetime \cite{Avelino:2012ge,Pani:2011mg,Sham:2013sya,Avelino:2012qe,Casanellas:2011kf}. When $0<\eta<\eta_-$, the BH has two event horizons located at $r_{H\pm}$
\begin{equation}
r_{H\pm}=\frac{\sqrt{  \left(M\pm\sqrt{M^2-Q^2}\right)^4-\eta Q^2}}{M\pm \sqrt{M^2-Q^2}}
\end{equation}
and when $\eta_-<\eta<\eta_+$ only one horizon at $r_{H+}$ survives.

\begin{figure}[htp!]
\centering
\includegraphics[width=0.45\textwidth]{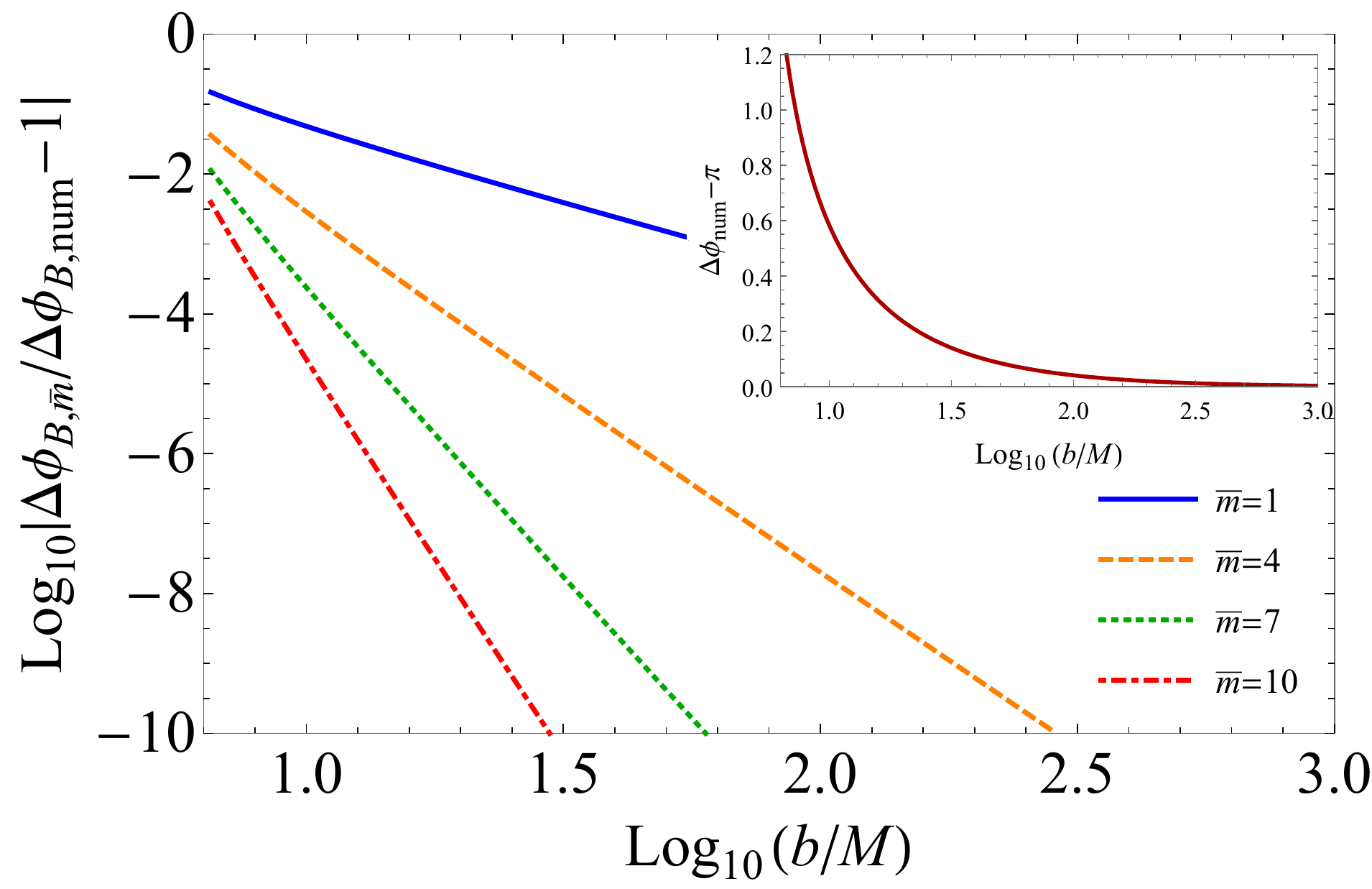}
\includegraphics[width=0.45\textwidth]{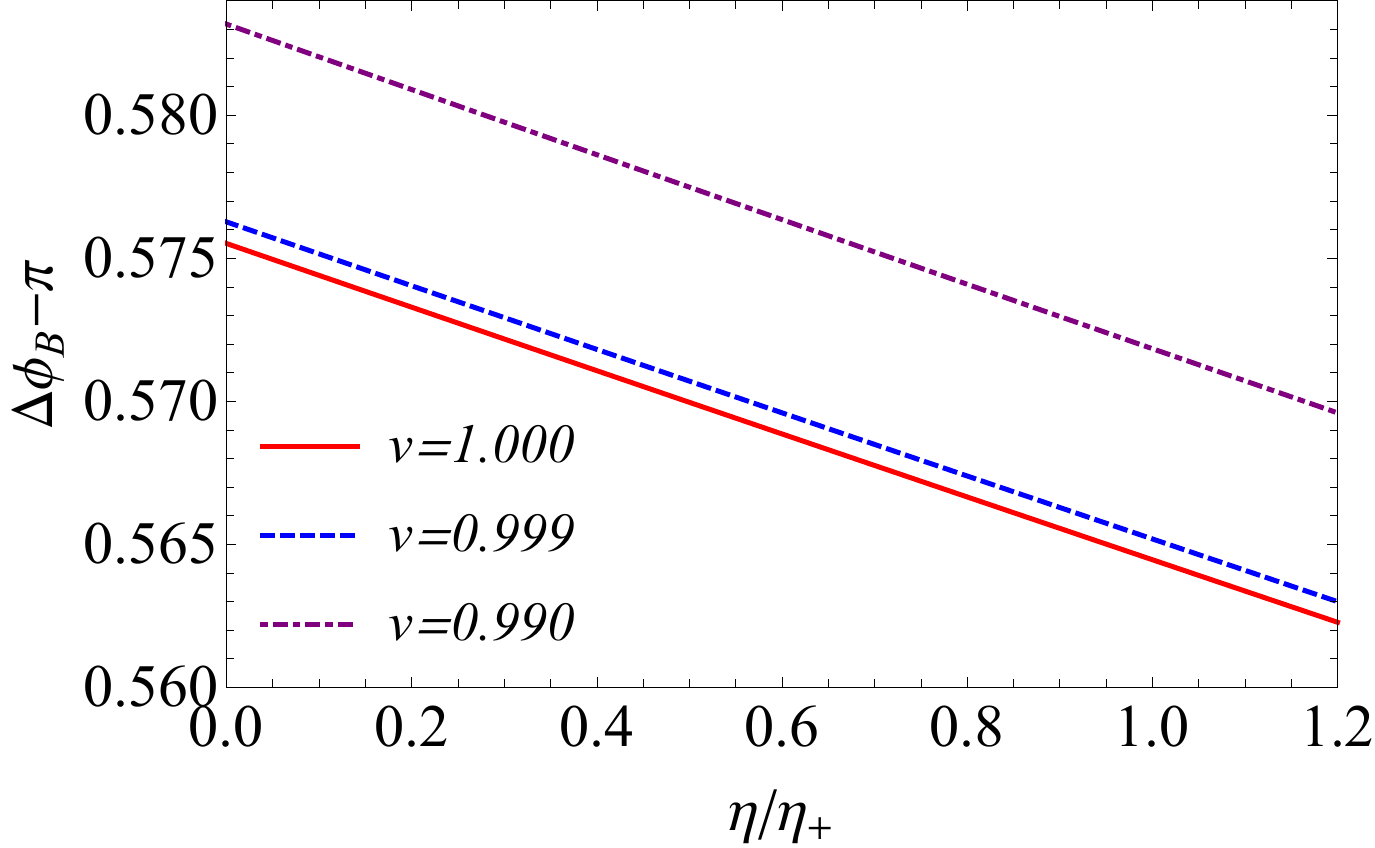}\\
(a) ~~~~~~~~~~~~~~~~~~~~~~~~~~~~~~~~~~~~~~~(b)
\caption{\label{fig:bifig1} (a) Difference between the perturbative $\Delta\phi_{B,\bar{m}}$ and the numerical result $\Delta\phi_{B,\mathrm{num}}$ as a function of $b/M$ from 7 to $10^3$. The inset shows the deflection angle itself for this range of $b$. Other parameters used are $Q=M/2,~v=1-10^{-2}, ~r_s=r_d=10^6M$ and $\eta=M/5$. (b) $\Delta \phi_B-\pi$ as a function of $\eta/M$ for  $b=10M$ and three values of $v=1,~0.999,~0.990$. Other parameters are the same as in (a).}
\end{figure}

Expanding metric functions \eqref{eq:bimetrics} at large $r$, we have
\begin{subequations}
\begin{align}
A(r)&=1- \frac{2M}{r}+ \frac{Q^2}{r^2}+ \frac{\eta Q^2}{r^4}+ \mathcal{O}\lb r\rb^{-5},\\
B(r)&=1+ \frac{2 M}{r}+\frac{4 M^2-Q^2}{r^2}+  \frac{8 M^3-4M Q^2}{r^3}+\frac{16M^4-12 M^2 Q^2+Q^4-2 \eta Q^2 }{r^4}+\mathcal{O}\lb r\rb^{-5},\\
\frac{C(r)}{r^2}&=1,
\end{align}
\end{subequations}
from which their first few coefficients $a_i,~b_i$ and $c_i$ are easily read off. Substituting these into Eq. \eqref{eq:yncoeffall}, the coefficients $y_{B,n}$ in this case are found to be
\begin{subequations}
\label{eq:ybi}
\begin{align}
y_{B,0}=&1,\\
y_{B,1}=&M  \left(1+\frac{1}{v^2}\right),\\
y_{B,2}=&M^2\left(\frac{3}{2}+\frac{6}{v^2}\right)-Q^2\left(\frac{1}{2}+\frac{1}{v^2}\right),\\
y_{B,3}=&M^2\left(\frac{5}{2}+\frac{45}{2v^2}+\frac{15}{2v^2}-\frac{1}{2v^6}\right)-Q^2\left(\frac{3}{2}+\frac{9}{v^2}+\frac{3}{2v^4}\right),\\
y_{B,4}=&M^4\left(\frac{35}{8}+\frac{70}{v^2}+\frac{70}{v^4}\right)-M^2Q^2\left(\frac{15}{4}+\frac{45}{v^2}+\frac{30}{v^4}\right)+Q^4\left(\frac{3}{8}+\frac{3}{v^2}+\frac{1}{v^4}\right)\nn\\
&-Q^2\eta\left(1+\frac{2}{v^2}\right). \label{eq:ybi4}
\end{align}
\end{subequations}
Substituting these into Eq. \eqref{eq:dphifinal}, the deflection in this Born-Infeld gravity becomes
\begin{align}
&\Delta \phi_{B}=\sum_{n=0}^{\infty} y_{B,n}\frac{ I_n(\theta_s,\theta_d)}{b^n}.
\label{eq:cacbi}
\end{align}
For null signals except photons, we can easily take the $v=1$ limit in Eq. \eqref{eq:ybi} to obtain its deflection angle.
Similar to the extended Einstein-Maxwell case in Eq. \eqref{eq:ynern4}, the parameter $\eta$ controlling the deviation from the RN spacetime also appears from the fourth order in Eq. \eqref{eq:ybi4}. We will see in Fig. \ref{fig:bifig1} that this similarity leads to quantitatively similar effect of the parameter $\eta$ on the deflection angle $\Delta\phi_B$ as that of the parameter $\alpha$ on the deflection angle $\Delta\phi_E$ in the extended Einstein-Maxwell gravity.

In Fig. \ref{fig:bifig1} (a), we plot the truncated deflection angles
\begin{equation}
\Delta \phi_{B,{\bar m}}=\sum_{n=0}^{\bar m}y_{B,n} \frac{ I_n(\theta_s,\theta_d)}{b^n},~(\bar{m}=1,2,\cdots)
\label{eq:cacbitrunc}
\end{equation}
to different orders as functions of the impact parameter.
The deflection angle $\Delta\phi_{B,\mathrm{num}}$ obtained by directly numerically integrating Eq. \eqref{eq:deltaphi} is compared with the truncated $\Delta\phi_{B,\bar{m}}$. It is seen that similar to the case of the extended Einstein-Maxwell spacetime, as the truncation order increases, the perturbative result approaches the numerical value for all $b$. Moreover, the larger the $b$ is, the smaller the difference between $\Delta\phi_{B,\bar{m}}$ and $\Delta\phi_{B,\mathrm{num}}$. When the truncation order $\bar{m}=10$, $\Delta\phi_{B,\mathrm{10}}$ is still a good approximation of the angle even for $b=10M$, at which point $\Delta\phi_B$ is also about 0.6 and the gravity is not that weak anymore.

With the perturbative result checked, we can
now use it to study the effect of spacetime parameter $\eta$ on the deflection angle.
In Fig. \ref{fig:bifig1} (b), we plotted $\Delta\phi_B$ as a function of $\eta$ from 0 to $1.2\eta_+$ for several $v$. It is seen that for the chosen values of other parameters, $\eta$ affects the deflection angle only by about 0.01 which is quite small comparing to the absolute value of the deflection angle itself. The effect of $\eta$ in this spacetime then is also much weaker than that of the charge $Q$ from $0$ to $M$, as was revealed in Ref. \cite{Pang:2018jpm}.

For photons, due to the influence of nonlinear electrodynamics, their motion can be regarded as moving along the geodesic line in an effective spacetime \cite{Jana:2015cha}. The metric functions of this spacetime is given by \begin{subequations}
\label{eq:bieffmetric}
\begin{align}
A_{\mathrm{eff}}(\bar{r})&=\left(1+\frac{\eta Q^2}{\bar{r}^4}\right) \left(1-\frac{2 M}{\bar{r}}+\frac{Q^2}{\bar{r}^2}\right),\\
B_{\mathrm{eff}}(\bar{r})&=\frac{1+\frac{\eta Q^2}{\bar{r}^4}}{1-\frac{2 M}{\bar{r}}+\frac{Q^2}{\bar{r}^2}},\\
C_{\mathrm{eff}}(\bar{r})&=\frac{\bar{r}^2 \left(1+\frac{\eta Q^2}{\bar{r}^4}\right)^2}{1-\frac{\eta Q^2}{\bar{r}^4}},
\end{align}
\end{subequations}
where $\bar{r}$ is related to $r$ by
\begin{equation}
\frac{\bar{r}^2 \left(1+\frac{\eta Q^2}{\bar{r}^4}\right)^2}{1-\frac{\eta Q^2}{\bar{r}^4}}=r^2.
\end{equation}
Clearly, the weak field limit of large $r$ is also the large $\bar{r}$ limit.

To compute the deflection angle of a photon in this metric, we first note that metric \eqref{eq:bieffmetric} has a conformal factor  $\left(1+\frac{\eta Q^2}{\bar{r}^4}\right)$ which will not affect the computation of photon's deflection angle (this can be recognized from Eq. \eqref{eq:deltaphi}). Therefore, we can remove this factor and deal with the reduced metric of the form
\begin{equation}
A_{\mathrm{red}}(\bar{r})=\frac{1}{B_{\mathrm{red}}(\bar{r})}=1-\frac{2 M}{\bar{r}}+\frac{Q^2}{\bar{r}^2}, \quad C_{\mathrm{red}}(\bar{r})=\frac{\bar{r}^2 \left(1+\frac{\eta Q^2}{\bar{r}^4}\right)}{1-\frac{\eta Q^2}{\bar{r}^4}}. \label{eq:bisimplifiedeffmetric}
\end{equation}
To use the perturbative method for this metric, we expand it at large $\bar{r}$
\begin{subequations}
\begin{align}
A_{\mathrm{red}}(\bar r)&=1-\frac{2M}{\bar r}+\frac{Q^2}{\bar{r}^2},\\
B_{\mathrm{red}}(\bar{r})&=1+ \frac{2 M}{\bar{r}}+\frac{4 M^2-Q^2}{\bar{r}^2}+  \frac{8 M^3-4M Q^2}{\bar{r}^3}+\frac{16M^4-12 M^2 Q^2+Q^4 }{\bar{r}^4}+\mathcal{O}\left(\bar{r}\right)^{-5},\\
\frac{C_{\mathrm{red}}(\bar{r})}{\bar{r}^2}&=1+\frac{2\eta Q^2}{\bar{r}^4}+\mathcal{O}\left(\bar{r}\right)^{-5}.
\end{align}
\end{subequations}
Substituting these into Eq. \eqref{eq:yncoeffall}, the first few coefficients in the deflection angle \eqref{eq:dphifinal} in this case are found to be
\begin{subequations}
\label{eq:ybip}
\begin{align}
y^{\prime}_{B,0}&=1,\\
y^{\prime}_{B,1}&=2M,\\
y^{\prime}_{B,2}&=\frac{15M^3}{2}-\frac{3Q^2}{2},\\
y^{\prime}_{B,3}&=32M^3-12M Q^2,\\
y^{\prime}_{B,4}&=\frac{1155M^4}{8}-\frac{315M^2 Q^2}{4}+\frac{35Q^4}{8}+3\eta Q^2. \label{eq:ybp4}
\end{align}
\end{subequations}
Comparing to the $v\to 1$ limit of the coefficients \eqref{eq:ybi} for other null signals (such as gravitational wavelets), we find that there is only one difference in the sign of $\eta$ in Eqs. \eqref{eq:ybi4} and \eqref{eq:ybp4}. A comparison in even high orders $y_{B,n}$ and $y^\prime_{B,n}~(n\geq 5)$
shows that their difference also only appear in the sign of $\eta$.
Therefore, we can conclude that from the asymptotic deflection angle point of view, the nonlinear parameter $\eta$ influences photon and other null signals with different signs. Note however, this does not mean the deflection angles change their signs because the effect of $\eta$ is only at the fourth order and above.

\subsection{Charged Ellis-Bronnikov spacetime\label{subsec:3}}

The charged Ellis-Bronnikov spacetime is the charged version of the Ellis-Bronnikov solution for the Einstein-Maxwell-(phantom)-dilaton theory.
The metric functions of this spacetime are \cite{Nozawa:2020wet}
\begin{subequations}
\label{eq:cebfunc}
\begin{align}
&A(r)=e^{-2\beta U}W^{2a_c^2/(a^2-a_c^2)},\\
&B(r)=\left\lbrack e^{2\beta U}W^{-2a_c^2/(a^2-a_c^2)}V\right\rbrack^{1/(n-3)}/V,\\
&C(r)=\left\lbrack e^{2\beta U}W^{-2a_c^2/(a^2-a_c^2)}V\right\rbrack^{1/(n-3)}r^2,
\end{align}
\end{subequations}
where $n$ is the dimension of the spacetime and
\begin{subequations}
\label{eq:cebpara}
\begin{align}
&U=\mathrm{arctan}\lb \frac{M}{2r^{n-3}}\rb,~W=\frac{1+q^2(a^2-a_c^2)e^{2\beta_\pm U}}{1+q^2(a^2-a_c^2)},~V=1+\frac{M^2}{4r^{2(n-3)}},\\
&a_c=\sqrt{\frac{2(n-3)}{n-2}},~\beta_\pm\equiv\pm\frac{a}{a_c}\sqrt{1+\beta^2}-\beta.
\end{align}
\end{subequations}
Here $q$ is the electric charge of the spacetime, $a$ is the dilaton coupling constant,  and $M,~\beta$ are parameters about the dilaton field in the Einstein-Maxwell-dilaton action. Note there is an equivalence between the solutions with parameters $(M,a,\beta)$ and $(-M,-a,-\beta)$. Therefore we can concentrate on the case $M>0$ but keep two branches of $\beta_\pm$. And for simplicity, we will set $n=4$ first so that $a_c=1$ and study the general $n$ case later.
This charged Ellis-Bronnikov spacetime with certain ranges of the parameters describes a traversable wormhole bridging two asymptotically flat regions. In this work, we will focus on the range $a>a_c=1$, which allows the existence of such wormhole.

The metric functions \eqref{eq:cebfunc} can be expanded for large $r$ to find their series forms
\begin{subequations}
\begin{align}
A(r)=&1-2Y\frac{M}{r}+\left(X^2+2Y^2\right)\frac{M^2}{r^2}+\left(\frac{Z}{3}X^3-2X^2Y-\frac{4}{3}Y^3+\frac{Y}{6}\right)\frac{M^2}{r^3}+\mathcal{O}\left(r\right)^{-4},\\
B(r)=&1+2Y\frac{M}{r}+\left(-X^2+2Y^2\right)\frac{M^2}{r^2}+\left(\frac{Z}{3}X^3-2X^2Y+\frac{4}{3}Y^3-\frac{Y}{6}\right)\frac{M^2}{r^3}+\mathcal{O}\left(r\right)^{-4},\\
\frac{C(r)}{r^2}=&1+2Y\frac{M}{r}+\left(-X^2+2Y^2+\frac{1}{4}\right)\frac{M^2}{r^2}+\left(\frac{Z}{3}X^3-2X^2Y+\frac{4}{3}Y^3+\frac{Y}{3}\right)\frac{M^2}{r^3}+\mathcal{O}\left(r\right)^{-4},
\end{align}
\end{subequations}
where to simplify the notation we have defined $X=\beta_\pm q/[(a^2-1)q^2+1]$, $Y=\beta/2-a_c^2qX$ and $Z=q(a^2-a_c^2)-1/q$.
Again, we can read off the coefficients $a_i,~b_i$ and $c_i$ and then use Eq. \eqref{eq:yncoeffall} to find the coefficients
\begin{subequations}
\label{eq:cec}
\begin{align}
y_{C,0}=&1,\\
y_{C,1}=&MY\left(1+\frac{1}{v^2}\right),\\
y_{C,2}=&M^2\left[-X^2\left(1+\frac{1}{v^2}\right)+Y^2\left(2+\frac{6}{v^2}\right)+\frac{1}{8}\right],\\
y_{C,3}=&M^3\left[\frac{Z}{2}X^3\left(1+\frac{1}{v^2}\right)-X^2Y\left(\frac{9}{2}+\frac{12}{v^2}+\frac{3}{2v^4}\right)\right.\nn\\
&\left.+Y^3\left(\frac{9}{2}+\frac{49}{2v^2}+\frac{15}{2v^4}-\frac{1}{2v^6}\right)+Y\left(\frac{1}{2}+\frac{1}{2v^2}\right)\right].
\end{align}
 \end{subequations}
Substituting them into Eq. \eqref{eq:dphifinal}, the deflection angle in the four-dimensional CEB spacetime is obtained as \be
\Delta\phi_C=\sum_{m=0}^\infty y_{C,m} \frac{I_m(\theta_s,\theta_d)}{b^n}. \label{eq:dphires102}
\ee
To be complete, we also give the infinite source and detector distance limit of this deflection
\begin{align}
\Delta\phi_C=&\pi+2Y\left(1+\frac{1}{v^2}\right)\frac{M}{b}+\frac{\pi }{2}\left[-X^2\left(1+\frac{1}{v^2}\right)+Y^2\left(2+\frac{6}{v^2}\right)+\frac{1}{8}\right]\frac{M^2}{b^2}\nn\\
&+\frac{4}{3}\left[\frac{Z}{2}X^3\left(1+\frac{1}{v^2}\right)-X^2Y\left(\frac{9}{2}+\frac{12}{v^2}+\frac{3}{2v^4}\right)\right.\nn\\
&\left.+Y^3\left(\frac{9}{2}+\frac{49}{2v^2}+\frac{15}{2v^4}-\frac{1}{2v^6}\right)+Y\left(\frac{1}{2}+\frac{1}{2v^2}\right)\right]\frac{M^3}{b^3}+\mathcal{O}\left(\frac{M}{b}\right)^4.
\end{align}

\begin{figure}[htp!]
\centering
\includegraphics[width=0.45\textwidth]{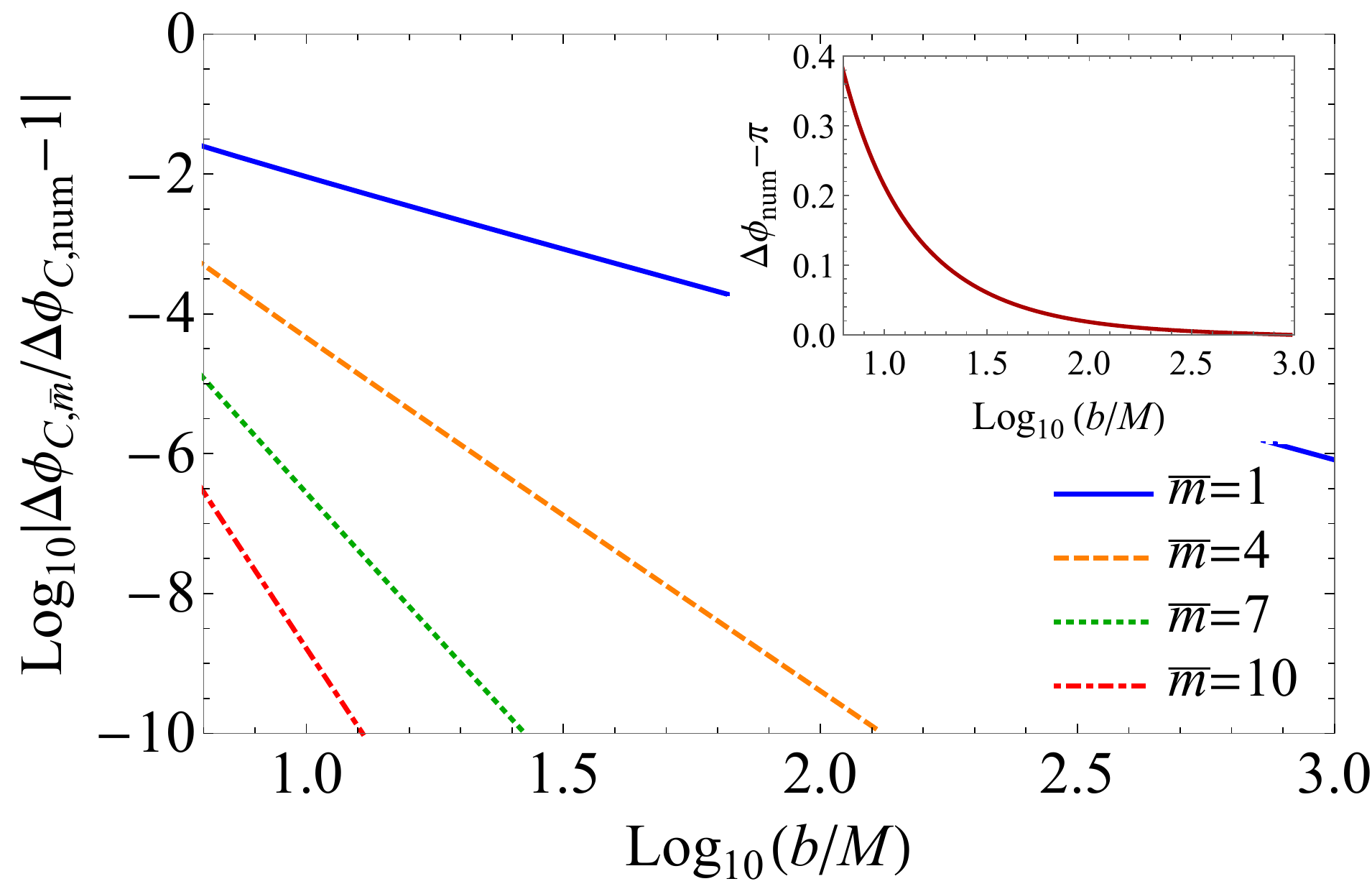}
\includegraphics[width=0.45\textwidth]{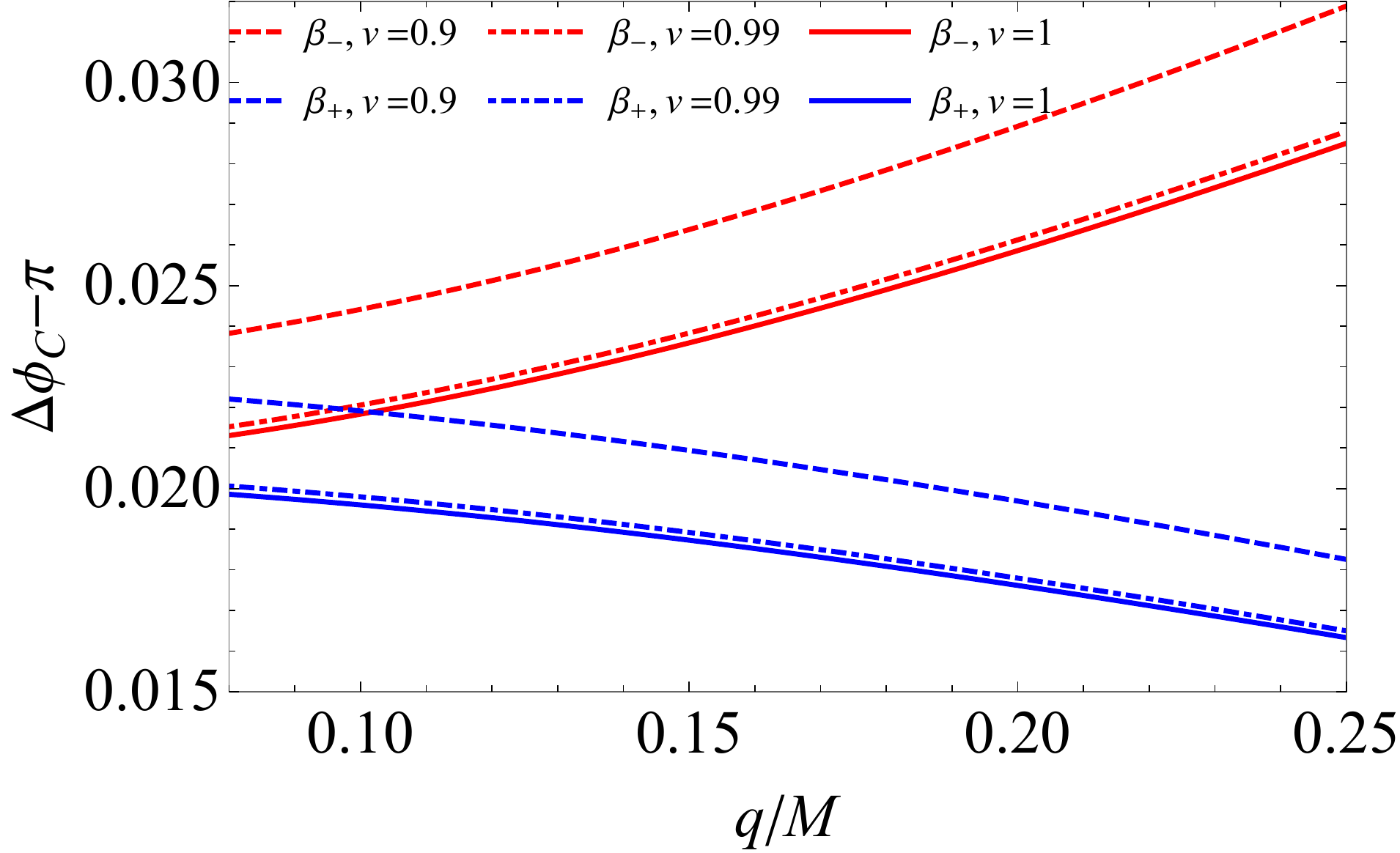}\\
(a)\hspace{0.4\textwidth}(b)\\
~\\
\includegraphics[width=0.45\textwidth]{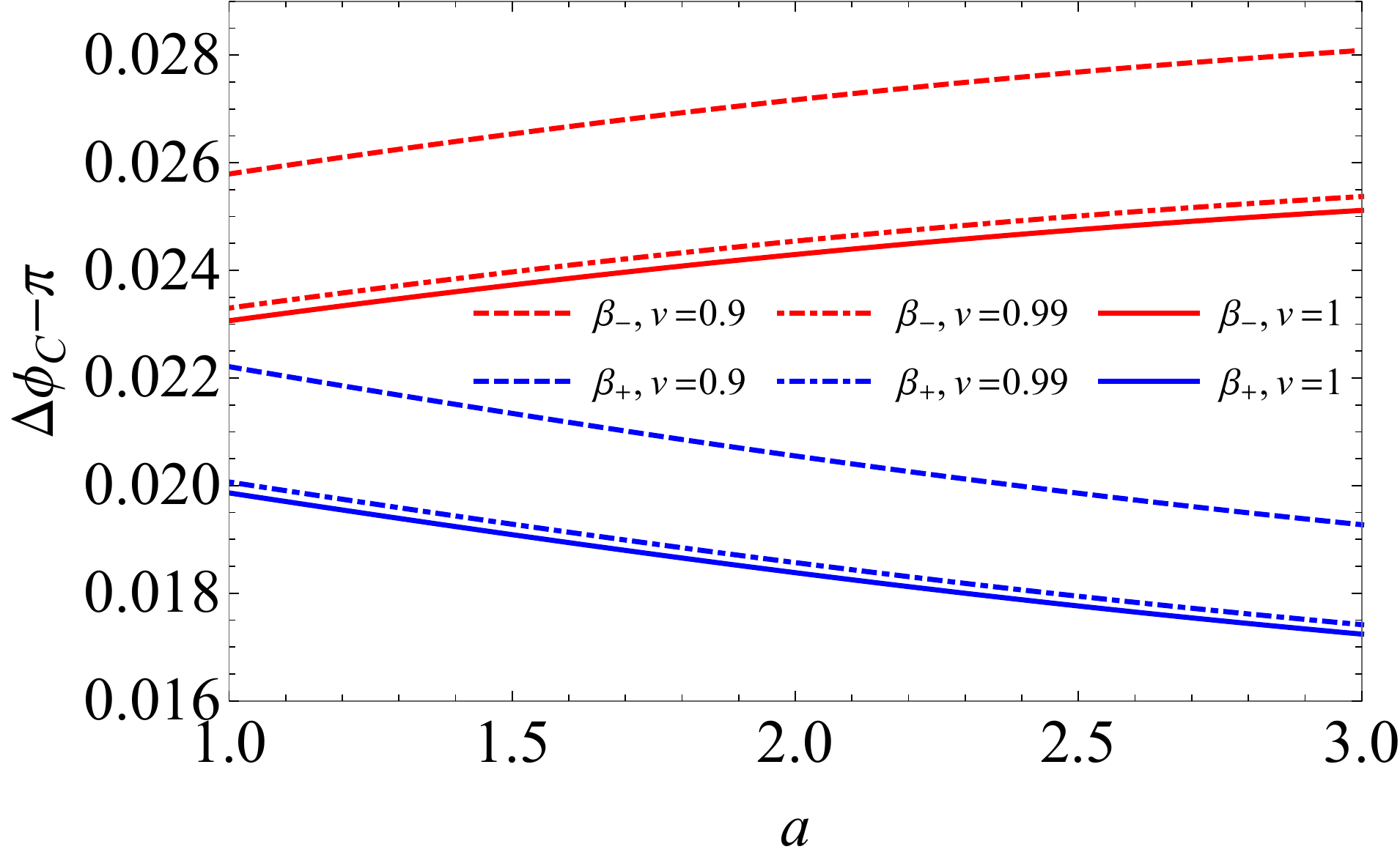}
\includegraphics[width=0.45\textwidth]{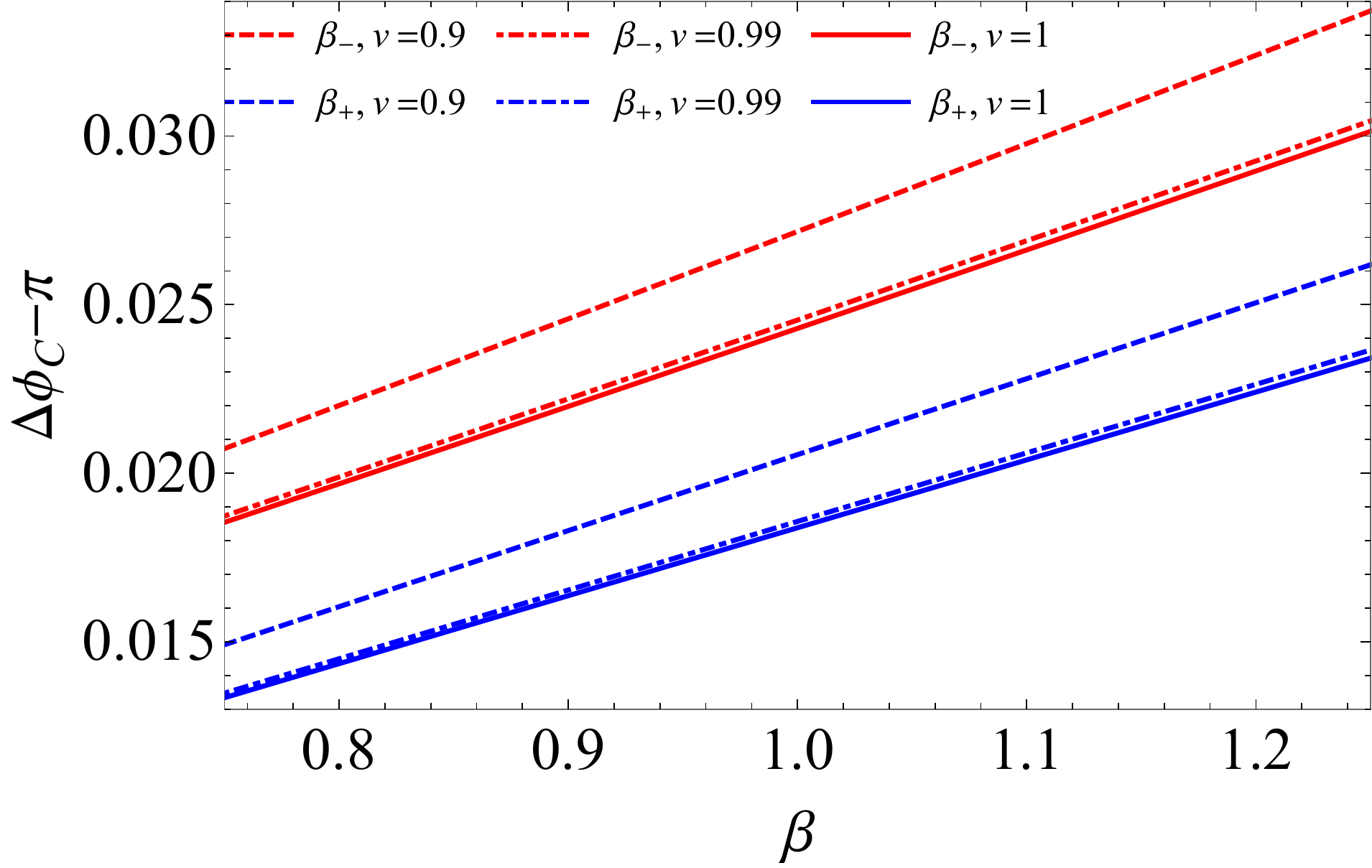}\\
(c)\hspace{0.4\textwidth}(d)\\
\caption{\label{fig:ceb} (a) Difference between the perturbative $\Delta\phi_{C,\bar{m}}$ and the numerical result $\Delta\phi_{C,\mathrm{num}}$ as a function of $b/M$ from 7 to $10^3$. The inset shows the deflection angle itself for this range of $b$. Other parameters used are $q=M/6,~\beta=1,~a=2,~v=1-10^{-2}, ~r_s=r_d=10^6M$ and we only showed for $\beta_+$. $\Delta \phi_C$ as a function of $q/M$ from $1/12$ to $1/4$  (b), of $a$ from $1$ to $3$ (c), and of $\beta$ from $3/4$ to $5/4$ (d). We plot for $b=100M$ and three values of $v=1,~0.99,~0.9$ in (b)-(d). Other parameters are the same as in (a).}
\end{figure}

In Fig. \ref{fig:ceb} (a), a comparison of the truncated deflection angle $\Delta\phi_{C,\bar{m}}$ and the numerical result is made for a range of $b$. It is seen that again, as in the previous two cases, the series result approaches the true physical value exponentially as the order increases. The difference between $\Delta \phi_{C,\bar{10}}$ and $\Delta \phi_{C,\mathrm{num}}$ is only about $10^{-6}$ even for $b=7M$ at which point the deflection is roughly 0.4. In Fig. \ref{fig:ceb} (b)-(d), the dependence of $\Delta\phi_C$ on the spacetime charge $q$, dilaton coupling constant $a$ and parameter $\beta$ are plotted. It is seen that as $q$ or $a$ increases, the deflection angles with branch $\beta_+$ (or $\beta_-$) will increase (or decrease) monotonically. On the other hand, both the deflections with branch $\beta_+$ and $\beta_-$ will increase as the parameter $\beta$ increases. Finally in all these plots, the deflection angles will decrease as the signal velocity $v$ increases, as expected.

In addition to the $n$=4 case, to examine the power of the perturbative method, we also investigated the deflection in the CEB spacetime with arbitrary spacetime dimension. Carrying out the asymptotic expansion of the metric \eqref{eq:cebfunc} with general $n$, it is found that \begin{subequations}
\begin{align}
A(r)=&1+\frac{2Y}{n-3}\frac{M}{r^{n-3}}+\left(a_c^2X^2+2Y^2\right)\frac{M^2}{r^{2n-6}}\nn\\
&-\left(\frac{a_c^2Z}{3} X^3+2a_c^2X^2Y+\frac{4}{3}Y^3-\frac{1}{6}Y\right)\frac{M^3}{r^{3n-9}}+\mathcal{O}\left(r\right)^{4(3-n)},\\
B(r)=&1-2Y\frac{M}{r^{n-3}}-\left[\frac{a_c^2X^2}{n-3}-\frac{2Y^2}{(n-3)^2}+\frac{n-4}{4}\right]\frac{M^2}{r^{2n-6}}\nn\\
&+\left[\frac{a_c^2Z X^3}{3(n-3)}-\frac{2a_c^2X^2Y^2}{(n-3)^2}+\frac{4Y^3}{3(n-3)^3}+\frac{n-9}{6(n-3)^2}\right]\frac{M^3}{r^{3n-9}}+\mathcal{O}\left(r\right)^{4(3-n)},\\
\frac{C(r)}{r^2}=&1-2Y\frac{M}{r^{n-3}}-\left[\frac{a_c^2X^2}{n-3}-\frac{2Y^2}{(n-3)^2}-\frac{1}{4(n-3)}\right]\frac{M^2}{r^{2n-6}}\nn\\
&+\left[\frac{a_c^2Z X^3}{3(n-3)}-\frac{2a_c^2X^2Y^2}{(n-3)^2}+\frac{4Y^3}{3(n-3)^3}-\frac{n-6}{6(n-3)^2}\right]\frac{M^3}{r^{3n-9}}+\mathcal{O}\left(r\right)^{4(3-n)}.
\end{align}
\end{subequations}
Reading off the coefficients $a_i,~b_i$ and $c_i~(i\in \mathbb{Z})$ and substituting into Eq. \eqref{eq:yncoeffall}, it is found that only the $m(n-3)~(m\in\mathbb{Z}_\geq)$ order coefficients $y_{C,m(n-3)}$ are nonzero. The first few of them are
\begin{subequations}
\begin{align}
y_{C,0}^\prime=&1,\\
y_{C,n-3}^\prime=&MY\left(1+\frac{n-3}{v^2}\right),\\
y_{C,2n-6}^\prime=&M^2\left\{-a_c^2X^2\left(1+\frac{n-3}{v^2}\right)+Y^2\left[2+\frac{6(n-3)}{v^2}+\frac{2(n-3)(n-4)}{v^4}\right]+\frac{1}{8}\right\},\\
y_{C,3n-9}^\prime=&M^3\left\{\frac{a_c^2Z}{2}X^3\left(1+\frac{n-3}{v^2}\right)-a_c^2X^2Y\left[\frac{9}{2}+\frac{12(n-3)}{v^2}+\frac{3(n-3)(3n-11)}{2v^4}\right]\right.\nn\\
&+Y^3\left[\frac{9}{2}+\frac{49(n-3)}{2v^2}+\frac{15(n-3)(3n-11)}{2v^4}+\frac{(n-3)(3n-11)(3n-13)}{2v^6}\right]\nn\\
&\left.+Y\left(\frac{1}{2}+\frac{n-3}{2v^2}\right)\right\}.
\end{align}
\end{subequations}
The deflection angle for general dimensional CEB spacetime is then found using Eq. \eqref{eq:dphifinal} as
\begin{align}
\Delta\phi_C^\prime=\sum_{m=0}^\infty y_{C,m(n-3)}^\prime \frac{I_{m(n-3)}(\theta_s,\theta_d)}{b^{m(n-3)}}. \label{eq:dphires11}
\end{align}

\begin{figure}[htp!]
\centering\includegraphics[width=0.45\textwidth]{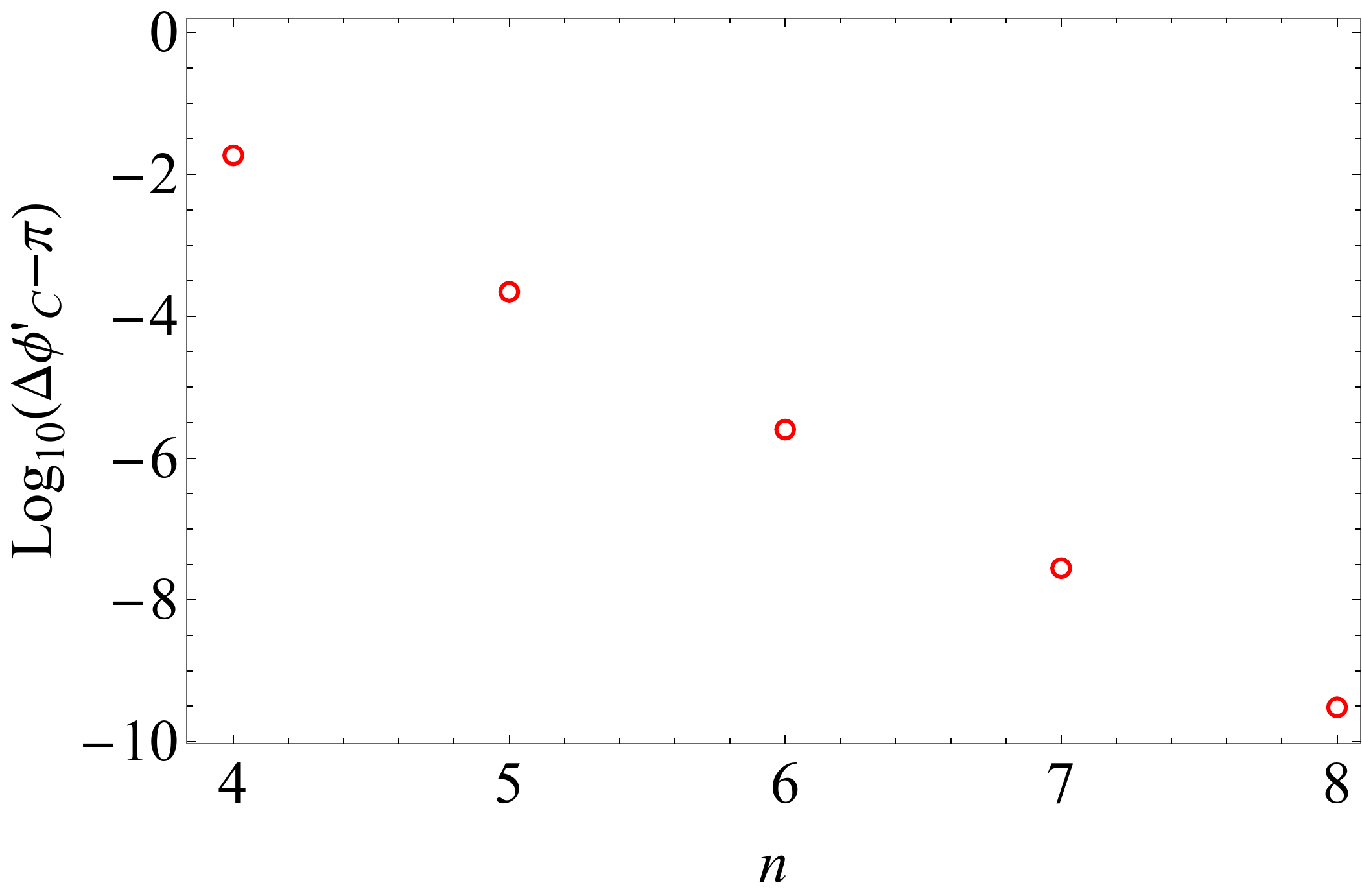}
\caption{\label{fig:nincrease} The deflection $\Delta\phi_C^\prime-\pi$ as a function of the spacetime dimension $n$. The parameters are the same as in Fig. \ref{fig:ceb} (a) and we chose $b=100M$. }
\end{figure}

From Eq. \eqref{eq:dphires11} we immediately recognize that the first nontrivial order of the deflection is about $ y_{C,n-3}^\prime /b^{n-3}$. Since $n>4$ and $b$ is usually larger than $M$ which characterize the spacetime, clearly the deflection angle will decrease roughly by a factor of $M/b$ when $n$ increases by $1$. In other words, the size of the deflection angle decreases roughly as a geometrical series with $M/b$ as the common ratio as $n$ increases. In Fig. \ref{fig:nincrease}, we plot the deflection angle \eqref{eq:dphires11} for $n=4,5,6,7,8$ for $b=100M$ and infinite $r_{s,d}$. Other parameters used in this figure are the same as in Fig. \ref{fig:ceb} (a). Clearly, as $n$ increases by 1 each time, the deflection decrease roughly by two orders of magnitude, which is exactly the $M/b$ value we took in the plot. We note that the dependence of $\Delta \phi_{C}$ on $n$ at the leading order as $\mathcal{O}(M/b)^{n-3}$ is also observed in Ref. \cite{Belhaj:2020rdb}.

Using this example, what we want to emphasize again is that if $r_{s,d}$ are finite, then their correction through $I_0(\theta_s,\theta_d)$ to $\Delta\phi_C^\prime$ will be much more important in higher dimensional spacetime than $n=4$. The reason is that the leading order finite distance correction, as revealed by Eq. \eqref{eq:dphiwithfd}, is always $-b/r_{s,d}$ regardless of the dimension $n$. However, in this example the first non-trivial order due to the impact parameter is proportional to $M^{n-3}/b^{n-3}$ which decrease rapidly as $n$ increases. Therefore, the higher the spacetime dimension, the more important the finite distance effect of the source and detector for this kind of spacetime.

\section{Extension to Asymptotically Non-flat spacetimes \label{sec:asynonm}}

In this section, we show that the perturbative method can be extended to some spacetimes that are asymptotically non-flat to find the deflection angle in them. We will take the BH solution in a nonlinear electrodynamical scalar theory as an example. Note that this spacetime is asymptotically non-flat for the spacetime parameters that we will study (see below Eq. \eqref{eq:abc0para}).

To extend the perturbative method to the asymptotically non-flat case, we will not explicitly repeat all the calculations but only point out the steps that are different from the original derivation in Sec. \ref{sec:method}. We will assume that the metric functions have asymptotic expansions of the form
\be
A(r) =\sum_{n=0} \frac{a_n}{r^n},\ B(r) =\sum_{n=0} \frac{b_n}{r^n},\ \frac{C(r)}{r^2}=\sum_{n=0} \frac{c_n}{r^n}, \label{eq:abcform2}
\ee
where $a_0,~b_0$ and $c_0$ are not all equal to one.

To use our method to this kind of spacetimes, we should first derive new relation between $(E,~L)$ and $(v,~b)$ from Eq. \eqref{eq:gdeq}. For a static observer at infinity, its four-velocity is given by  $Z^\mu=\left(1/\sqrt{a_0},~0,~0,~0\right)$.
A particle with four-velocity $U^\mu=\left(\dd t/\dd\tau,~\dd r/\dd\tau,~0,~\dd\phi/\dd\tau\right)$ then will be detected to have velocity
\be
v^\mu= h^\mu_\nu U^\nu/\gamma=\frac{1}{\gamma}\left(0,~\frac{\dd r}{\dd\tau},~0,~\frac{\dd\phi}{\dd\tau}\right),
\label{eq:fvdef}
\ee
in this observer's inertial coordinate system. Here $h_{\mu\nu}=g_{\mu\nu}+Z_\mu Z_\nu$ is the induced metric and
\begin{align}
\gamma=-Z^\mu U_\mu=\sqrt{a_0}\dd t/\dd\tau \label{eq:gammadef}
\end{align}
is the $\gamma$ factor. The asymptotic velocity $v$ of this signal is defined as the norm of $v^\mu$
\begin{align}
v^2=h_{\mu\nu}v^\mu v^\nu\big|_{r\to\infty}=
\frac{1}{\gamma^2}\lsb b_0\left(\frac{\dd r}{\dd\tau}\rb^2+c_0r^2\lb \frac{\dd\phi}{\dd\tau}\right)^2\rsb.\label{eq:vsqdef}
\end{align}
The proper time of the signal at infinity satisfies the following relation
\begin{align}
-\dd \tau^2&=\dd s^2=-a_0\dd t^2+b_0\dd r^2+c_0r^2\dd\phi^2\\
&=-a_0\dd t^2+\gamma^2v^2\dd\tau^2\\
&=-a_0(1-v^2)\dd t^2,
\end{align}
where in the second and third step Eqs.  \eqref{eq:vsqdef} and \eqref{eq:gammadef} are used respectively.
From this, we are able to obtain that for the static observer at infinity
\begin{align}
\frac{\dd t}{\dd \tau}=\frac{1}{\sqrt{a_0(1-v^2)}},~\gamma=\frac1{\sqrt{1-v^2}}. \label{eq:dtodtaugamma}
\end{align}
Combining this with Eq. \eqref{eq:gdeq1}, we found the new relation between $E$ and $v$
\begin{align}
E=A(r)\frac{\dd t}{d\tau}\Big|_{r\to\infty}=\frac{\sqrt{a_0}}{\sqrt{1-v^2}}. \label{eq:enew}
\end{align}

To get a proper definition of the impact parameter $b$ in this case, we recall that its geometric definition is the distance from the lens center to the straight asymptotic line of the trajectory, and therefore
\begin{align}
b\equiv \sin\delta\cdot d_{LO}\big|_{r\to\infty}\approx \frac{|{\bf v}_\phi|}{v} d_{OL}\big|_{r\to\infty}
=\frac{v^\phi d_{OL}}{v} d_{OL}\big|_{r\to\infty}.
\end{align}
Here $\delta$ is the angle by the three velocity ${\bf v}$ itself and its $\phi$-component ${\bf v}_\phi$, and $d_{OL}$ is the distance from the observer to the lens.
Using \eqref{eq:fvdef} for $v^\phi$, Eq. \eqref{eq:dtodtaugamma} for $\gamma$ and $d_{OL}=\sqrt{c_0} r$, this becomes
\begin{align}
b=c_0r^2\frac{\sqrt{1-v^2}}{v}\frac{\dd\phi}{\dd \tau} .
\end{align}
Finally substituting this equation into Eq. \eqref{eq:gdeq2}, we obtain the new relation between $L$ and $(b,~v)$ as
\begin{equation}
L=C\frac{\dd\phi}{\dd\tau}\Big|_{r\to\infty}=\frac{bv}{\sqrt{1-v^2}}. \label{eq:lnew}
\end{equation}
It is seen that by comparing to Eq. \eqref{eq:leandbv} the relation between $L$ and $(b,~v)$ is not affected by the asymptotic non-flatnesss of the metric.

Corresponding to the new relations \eqref{eq:enew} and \eqref{eq:lnew}, the function $p(x)$ in Eq. \eqref{eq:pfuncdef} should also be revised to
\bea
\frac{1}{b}&=&\frac{1}{\sqrt{a_0}}\frac{\sqrt{E^2-a_0\kappa}}{\sqrt{E^2-\kappa A(r_0)}}\sqrt{\frac{A(r_0)}{C(r_0)}}\equiv p\lb \frac{1}{r_0}\rb. \label{eq:pfuncdefnew}
\eea
With these changes, formally $\Delta\phi$  still take the form of Eq. \eqref{eq:dphifinal}, i.e.,
\be
\Delta\phi =\sum_{n=0}^\infty y_n\frac{I_n(\theta_s,\theta_d)}{b^n} \label{eq:dphifinal2}
\ee
but with the following revised series coefficients
\begin{subequations}
\label{eq:ynnew}
\begin{align}
y_0=&\sqrt{\frac{b_0}{c_0}},\label{eq:y0coeff}\\
y_1=&\sqrt{\frac{b_0}{c_0}}\left(\frac{b_1\sqrt{c_0}}{2b_0}\right)\left(1-\frac{a_1b_0}{a_0b_1}\frac{1}{v^2}\right),  \\
y_2=&\sqrt{\frac{b_0}{c_0}}\left(\frac{b_1\sqrt{c_0}}{2b_0}\right)^2\left[\frac{2 b_0^2 c_2}{b_1^2 c_0}-\frac{b_0^2 c_1^2}{2 b_1^2 c_0^2}+\frac{b_0 c_1}{b_1 c_0}+\frac{2 b_0 b_2}{b_1^2}-\frac{1}{2}\nn\right.\\
&\left.-\left(\frac{2 a_1 b_0^2 c_1}{a_0 b_1^2 c_0}+\frac{4 a_2 b_0^2}{a_0 b_1^2}-\frac{4 a_1^2 b_0^2}{a_0^2 b_1^2}+\frac{2 a_1 b_0}{a_0 b_1}\right)\frac{1}{v^2}\right].
\end{align}
\end{subequations}
One can easily check that this system reduces to Eq. \eqref{eq:yncoeffall} when $a_0=b_0=c_0=1$.
In Eq. \eqref{eq:dphifinal2}, the $I_n$ are still given by Eq. \eqref{eq:inres2} while the $\theta_s$ and $\theta_d$ are still given by Eq. \eqref{eq:thetasddef} with new function $p(x)$ given by Eq. \eqref{eq:pfuncdefnew}.

\subsection*{Application to NES theory}

The metric functions of a nonlinear electrodynamical massless scalar theory are given by \cite{Tahamtan:2020mbb}
\begin{align}
A(r)=\frac{1}{B(r)}=-\frac{1}{\sqrt{2}\beta}\lsb \frac{r-r_1}{r-r_2}\rsb^\nu,~C(r)=\beta^2(r-r_1)(r-r_2)\lsb\frac{r-r_2}{r-r_1}\rsb^\nu, \label{eq:neabc}
\end{align}
where $\beta=q_m-\sqrt{2}$ is the electrodynamical strength parameter with $q_m$ being a non-negative magnetic charge. Since $\beta$ has to be negative in order for the temporal metric function to have the correct sign, this effectively set a range for $q_m$ to be $0\leqslant q_m<\sqrt{2}$. The parameter $\nu\in (0,1)$ in Eq. \eqref{eq:neabc} characterizes the strength of the scalar field. $r_1>r_2$ are two constants appearing in the NES theory solution, and it was shown that $r=r_1$ is the location of the event horizon of the BH in this spacetime \cite{Tahamtan:2020mbb}. Moreover, unlike event horizons of Schwarzschild or RN BHs, this event horizon is a true physical singularity of the spacetime. Note that this spacetime reduces to the Janis-Newmann-Winicour spacetime in the limit $\beta\to -\sqrt{2}$ and $r_1\to -r_2$ \cite{Tahamtan:2020mbb}.

The metric \eqref{eq:neabc} can be expanded asymptotically as
\begin{subequations}
\label{eq:metricexp}
\begin{align}
A(r)=&-\frac{1}{\sqrt{2} \beta }\left[1- \frac{\nu  \left(r_1-r_2\right)}{r}+\frac{\nu  \left(r_1-r_2\right) \left(\nu  r_1-\nu  r_2-r_1-r_2\right)}{2 }\frac{1}{r^2}\right]+\mathcal{O}\left(r\right)^{-3},\label{eq:arexpnes}\\
B(r)=&-\sqrt{2} \beta\left[1+ \frac{\nu  \left(r_1-r_2\right)}{r}+\frac{  \nu  \left(r_1-r_2\right) \left(\nu  r_1-\nu  r_2+r_1+r_2\right)}{2}\frac{1}{r^2}\right]+\mathcal{O}\left(r\right)^{-3},\\
\frac{C(r)}{r^2}=&\beta ^2\left\{1+ \left[(\nu -1) r_1-(\nu +1) r_2\right]\frac{1}{r}\nn\right.\\
&\left.+  \left[- \left(\nu ^2-1\right) r_2 r_1+\frac{1}{2}\nu\left(\nu -1\right)  r_1^2+\frac{1}{2}\nu  (\nu +1) r_2^2\right]\frac{1}{r^2}\right\}+\mathcal{O}\left(r\right)^{-3}.
\end{align}
\end{subequations}
From this, one can immediately see that at infinity
\begin{align}
a_0=\frac{1}{b_0}=-\frac{1}{\sqrt{2} \beta}, ~c_0=\beta^2.\label{eq:abc0para}
\end{align}
If $\beta$ were $-\sqrt{2}$, i.e., $q_m$ were zero, then a scaling  $t'=t/\sqrt{2},~r'=\sqrt{2}r$ will be able to transform the NES solution to an asymptotically flat spacetime, which we will not be interested in this work. When $q_m$ is nonzero and $\beta\neq -\sqrt{2}$ however, we can compute the Ricci and Riemann tensors of the metric and find that there exist components that will not asymptotically vanish. For example, asymptotically
\be
R_{\theta\theta}=\lb 1+\frac{\beta}{\sqrt{2}}\rb -\frac{(r_1-r_2)\beta \nu}{\sqrt{2}r}+\mathcal{O}\lb \frac{1}{r}\rb^2=R_{\theta\phi\theta}{}^\phi
\ee
and therefore the spacetime is asymptotically non-flat.

Reading off the coefficients $a_i,~b_i$ and $c_i$ from Eq. \eqref{eq:metricexp} and substituting into Eq. \eqref{eq:ynnew}, we are able to obtain
\begin{subequations}
\label{eq:ynnes}
\begin{align}
y_{N,0}&=\frac{\sqrt[4]{2}}{\sqrt{-\beta }},\\
y_{N,1}&=\frac{\sqrt[4]{2}}{\sqrt{-\beta }}(-\beta M)\left(1+\frac{1}{v^2}\right),\\
y_{N,2}&=\frac{\sqrt[4]{2}}{\sqrt{-\beta }}(-\beta M)^2\left(2-\frac{1}{2 \nu ^2}+\frac{6}{v^2}\right),
\end{align}
\end{subequations}
where we have defined an effective ADM mass $M=\nu(r_1-r_2)/2$ of the system, inspired by the first order of Eq. \eqref{eq:arexpnes}.
Substituting these into Eq. \eqref{eq:dphifinal2}, we are able to obtain the deflection angle $\Delta\phi_N$ in this NES spacetime
\be
\Delta\phi_N=\sum_{n=0}^\infty y_{N,n} \frac{I_n(\theta_s,\theta_d)}{b^n}, \label{eq:dphires104}
\ee
where $I_n$ are still given by Eq. \eqref{eq:inres2} while the $\theta_s$ and $\theta_d$ are still given by Eq. \eqref{eq:thetasddef} with new function $p(x)$ given by Eq. \eqref{eq:pfuncdefnew} for the metric \eqref{eq:neabc}.
From the coefficients $y_{N,n}$ in Eq. \eqref{eq:ynnes} and the higher order ones, we can tell that in order for Eq. \eqref{eq:dphires104} to converge, the ratio of the adjacent terms in this series has to be smaller than one. When $\nu$ is small, inspecting Eq. \eqref{eq:ynnes} shows that this condition effectively becomes $-\beta M/(\nu^2 b)\lesssim 1$, i.e., $\nu>\sqrt{-\beta M/b}=\sqrt{-(q_m-\sqrt2)M/b}$, which can be thought as a condition for the deflection angle \eqref{eq:dphires104} to be valid.
In the $r_s,r_d\to\infty$ limit, $\Delta\phi_N$
becomes much simpler
\begin{align}
\Delta\phi_N=\frac{\sqrt[4]{2}}{\sqrt{-\beta }}\left[\pi-2\beta M\left(1+\frac{1}{v^2}\right)\frac{1}{b}+\pi\beta^2 M^2 \left(1-\frac{1}{\nu^2}+\frac{3}{v^2}\right)\frac{1}{b^2}\right]+\mathcal{O}\left(\frac{M}{b}\right)^3.\label{eq:nesinf}
\end{align}

\begin{figure}[htp!]
\centering
\includegraphics[width=0.45\textwidth]{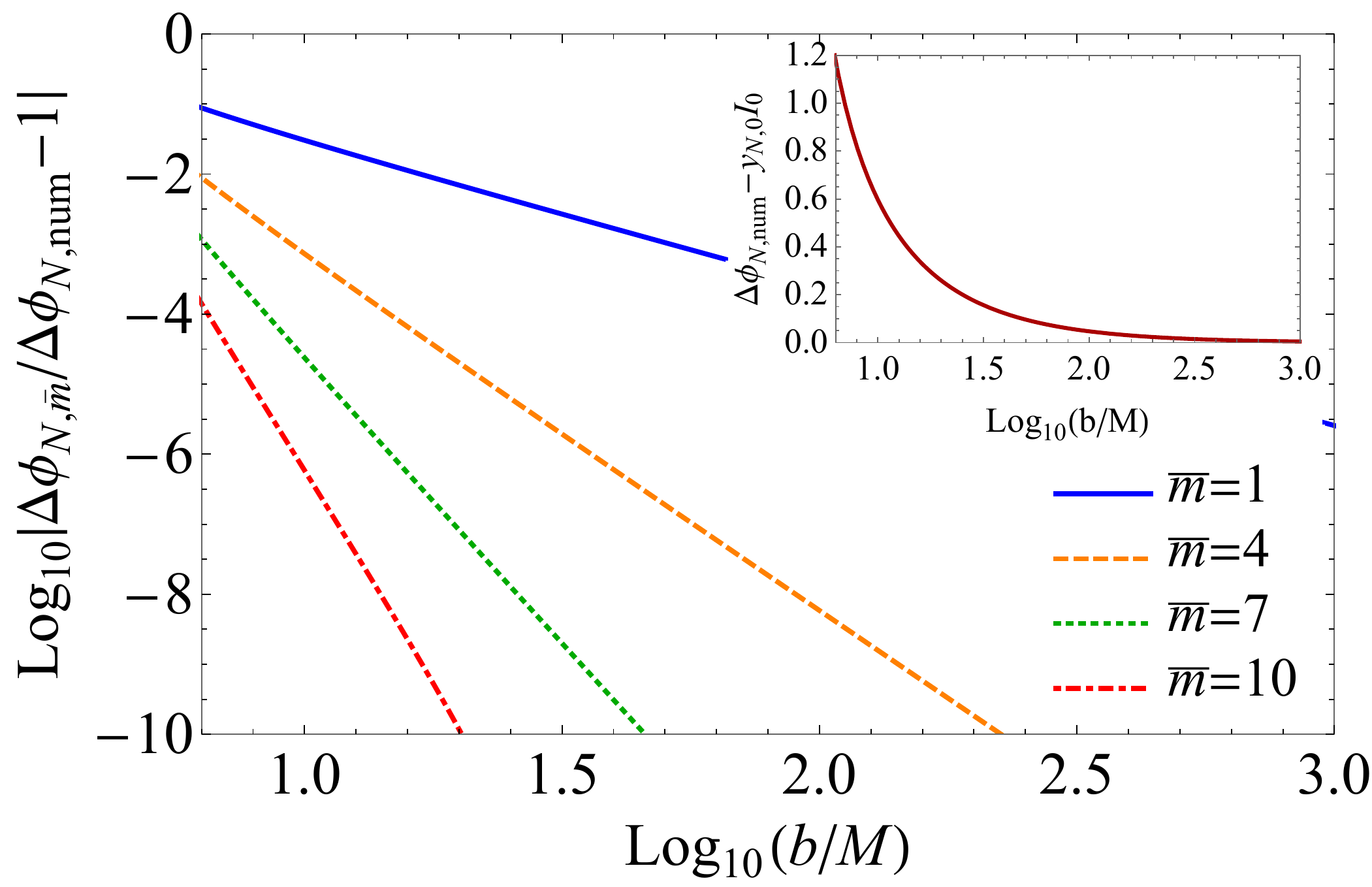}~
\includegraphics[width=0.45\textwidth]{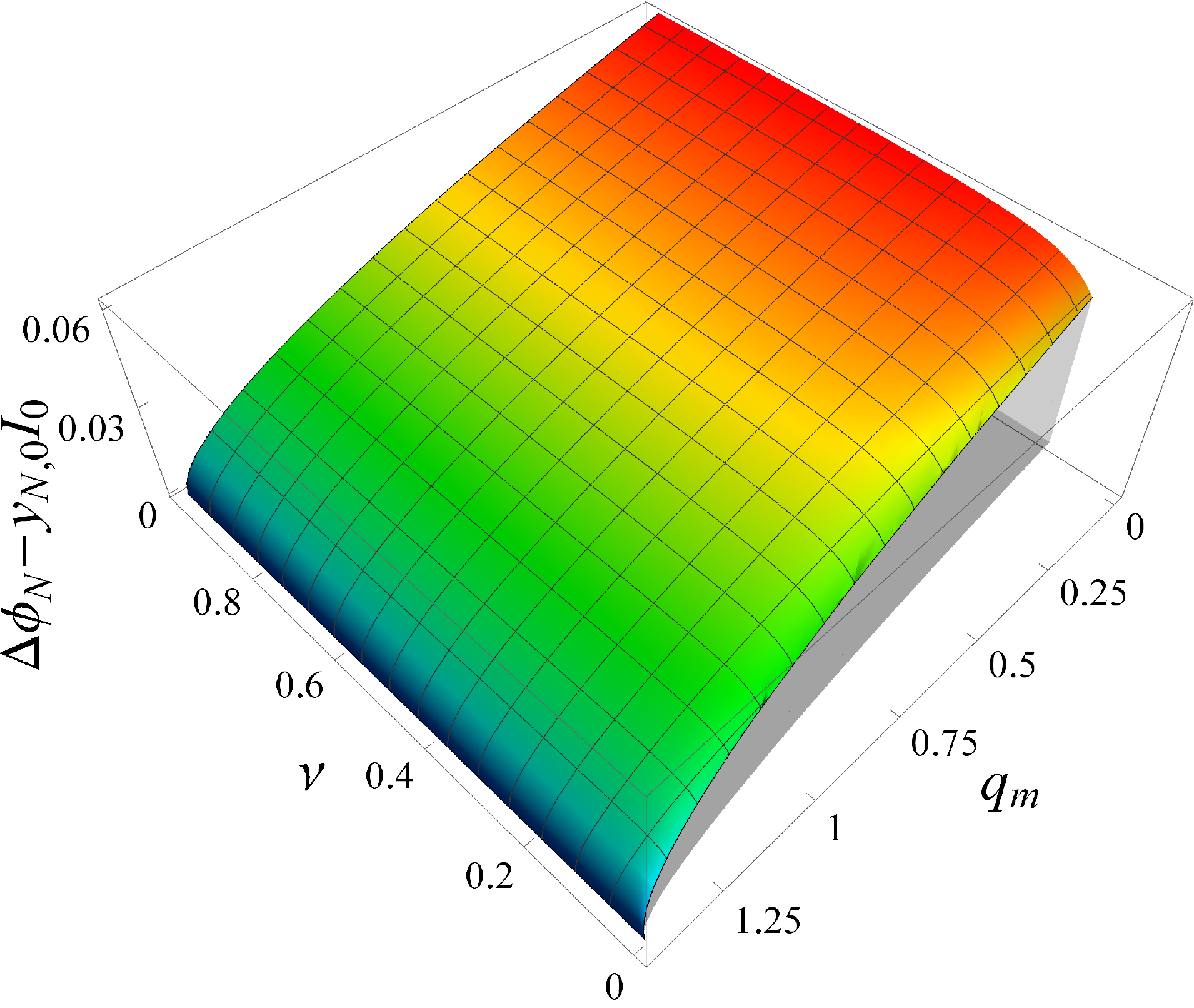}\\
(a)~~~~~~~~~~~~~~~~~~~~~~~~~~~~~(b)
\caption{(a) Difference between the perturbative $\Delta\phi_{N, \bar{m}}$ and the numerical result $\Delta\phi_{N,\mathrm{num}}$ as a function of $b/M$ from 7 to $10^3$. The inset shows the deflection angle itself for this range of $b$. Other parameters used are $q_m=1/2,~v=1-10^{-2}, ~r_s=r_d=10^6M$ and $\nu=1/2,~r_1=-r_2=M/\nu$. (b) $\Delta \phi_N$ as a function of $q_m$ and $\nu$ for $b=100M$. Other parameters are the same as in (a).
\label{fig:nes}}
\end{figure}

Similar to the cases in Sec. \ref{sec:applications}, we can define a truncated partial sum $\Delta\phi_{N,\bar{m}}$ and compare it with the numerical integration result, as shown in Fig. \ref{fig:nes}(a). Again, this shows that the perturbative result works very well, even for smaller $b$ at which the deflection angle is not that small anymore.
While in Fig. \ref{fig:nes} (b), we plot the dependence of $\Delta\phi_N$ on its parameters $0\leqslant q_m<\sqrt{2}$ and $\sqrt{(\sqrt{2}-q_m)M/b}<\nu<1$ for $b=100M$. It is seen that for a fixed $\nu$, as the magnetic charge increases, the deflection angle decreases. This is qualitatively similar to the effect of electrical charge in some charged spacetimes, such as the RN spacetime \cite{Pang:2018jpm,Jia:2020xbc}. For fixed $q_m$ however, as $\nu$ increases, the deflection angle keeps increasing, although the increase rate becomes very small after about $0.2$.

Theoretically, we observe from Eq. \eqref{eq:y0coeff} (or Eq. \eqref{eq:nesinf} in the NES spacetime)
one of the most fundamental feature for geodesic motion in an asymptotically non-flat spacetime whose metrics are like Eq. \eqref{eq:abcform2} with $c_0/b_0\neq 1$ or \eqref{eq:metricexp} with $\beta\neq -\sqrt{2}$: the deflection angle at leading order is not $\pi$ anymore. Since $-\sqrt{2}<\beta<0$, using Eqs. \eqref{eq:y0coeff} and \eqref{eq:metricexp}, we have $y_0=\sqrt{b_0/c_0}=\sqrt[4]{2}/\sqrt{-\beta}>1$. Combining with $I_0(r_i
\to\infty)=\pi/2$, this means that the deflection angle with an infinite $b$ in this spacetime is larger than $\pi$.

\begin{figure}[htp!]
\centering
\includegraphics[width=0.45\textwidth]{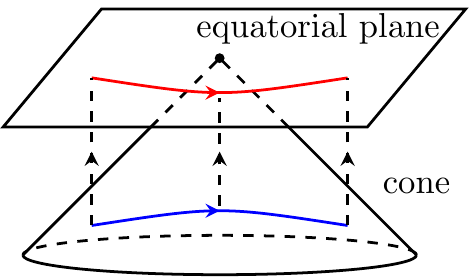}~~~
\includegraphics[width=0.3\textwidth]{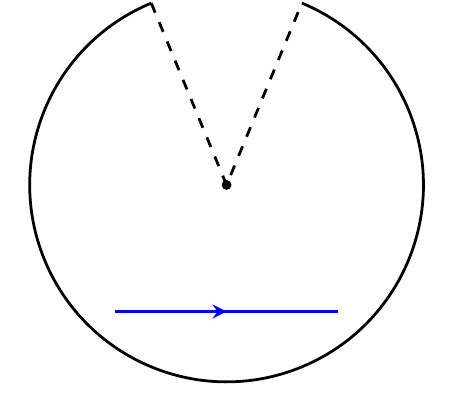}\\
(a)\hspace{7cm}(b)
\caption{(a) The trajectory in the equatorial plane and the cone which project to the plane. (b) The unwrapped cone and the geodesic on its surface.
\label{fig:geomexp}}
\end{figure}

The above feature for such kind of the asymptotically non-flat spacetimes actually can be understood in the following geometrical way.
First we notice that asymptotically, the metric for the equatorial plane of the spacetime \eqref{eq:neabc} can be approximated by
\begin{align}
\dd s^2=-a_0\dd t^2+b_0\dd r^2+c_0r^2\dd \phi^2,~b_0>c_0>0.
\label{eq:metriceqp}\end{align}
Now consider a cone structure
\begin{align}
\frac{z^2}{r^2}=\frac{b_0}{c_0}-1>0
\label{eq:zandr}
\end{align}
living in a spacetime with line element
\begin{align}
\dd s^2=-a_0\dd t^2+c_0\lsb \dd r^2+r^2\dd \phi^2+\dd z^2\rsb,\label{eq:conem}
\end{align}
then substituting Eq. \eqref{eq:zandr} into Eq. \eqref{eq:conem} one sees that the induced metric of the cone will be exactly the same as Eq. \eqref{eq:metriceqp}.
Therefore, any geodesics on the cone is also a geodesic on the equatorial plane after simple projection $(r,\phi,z)\to(r,\phi)$, as illustrated in Fig. \ref{fig:geomexp} (left).

On the other hand, it is known that a cone, after cut along its generatrix, is flat, and therefore any geodesic on it is necessarily a straight line. However, when wrapped into a cone and projected onto the equatorial plane, apparently its deflection angle at the leading order will be larger than $\pi$, as seen from Fig. \ref{fig:geomexp} (right). Therefore this simple geometric analogy explains the result in the leading order of Eq. \eqref{eq:nesinf}.

Lastly we also want to point out that although the metrics like Eq. \eqref{eq:abcform2} with $c_0/b_0\neq 1$ are asymptotically non-flat, not all asymptotically non-flat metrics take that form.
For example, typical asymptotically de-Sitter spacetimes are also asymptotically non-flat because of the extra $\Lambda r^2$ term in the metric functions $A(r)$ and $B(r)$. However such spacetimes still have a deflection angle at lowest order to be $\pi$ \cite{Li:2021qei,Liu:2022hbp,Parbin:2022iwt}. In other words, asymptotic non-flatness is only a necessary but not sufficient condition for the bending angle at the lowest order to be non-$\pi$. On the opposite side, we do are able to claim that when the metric is asymptotically non-flat in the way prescribed by Eq. \eqref{eq:abcform2} with $c_0/b_0\neq 1$, the deflection at the lowest order will not be $\pi$.

\section{Conclusions\label{sec:conc}}

In this work, we used a perturbative technique to compute the deflection of null and timelike signals in arbitrary static and spherically symmetric spacetimes of arbitrarily high dimension in the weak field limit. The technique naturally takes into account the finite distance effect of the source and observer. The resultant deflection angle takes a (quasi-)power series form of $M/b$, which can be converted to a dual-power series form of $M/b$ and $b/r_{s,d}$. We then applied this method to the extended Einstein-Maxwell spacetime, the Born-Infeld gravity and the charged Ellis-Bronnikov spacetime. It is shown by comparison to numerical method that the perturbative deflection angle works extremely well. In particular, in the last spacetime, the deflection at the leading non-trivial order is found to  decrease as $\mathcal{O}(M/b)^{n-3}$ as the dimension $n$ increases.

In the second part, we extended the perturbative method to some asymptotically non-flat spacetimes and applied the result to a nonlinear electrodynamical scalar theory. It is shown that one fundamental feature of this kind of spacetimes is
that the deflection angle at the leading order is not $\pi$ anymore. We provided a simple geometrical understanding of this feature.
In each of the above spacetimes, we studied the effects of the spacetime parameters and the signal velocity on the deflection angle.

Regarding possible further development of this work, there are at least the following possibilities. The first is to use the deflection angles found in these spacetimes to study the gravitational lensing effect, including the time delay between lensed images. The second is to further develop the method to handle other types of spacetimes such as those with logarithmic asymptotic expansions \cite{Li:2012zx,Zhang:2020mxi}. We are currently working along these directions.

\acknowledgements
We thank Haotian Liu for helping with the schematic diagrams. This work is supported by (China-Ukraine IGSCP-12).
\appendix

\section{Integration formulas \label{sec:appd}}

This integral \eqref{eq:indef} can be carried out using a change of variable $u=\sin\xi$ to find an elementary expression for even and odd non-negative $m$ respectively \cite{bk:inttable}
\begin{align}
&I_m(\theta_s,~\theta_d)=\lsb \int_{\theta_s}^{\pi/2}+\int_{\theta_d}^{\pi/2}\rsb \sin^m\xi\dd \xi ~(m\in\mathbb{Z}_\geq)\\
=&\sum_{i=s,d} \frac{(m-1)!!}{m!!}\times
\begin{cases}
\displaystyle  \left(\frac{\pi}{2}-\theta_i
+\cos\theta_i\sum_{j=1}^{[\frac{m-1}{2}]} \frac{(2j-2)!!} {(2j-1)!!}\sin^{2j-1} \theta_i\right),~m~\text{is even},\\
\displaystyle  \cos\theta_i \left(1
+\sum_{j=1}^{[\frac{m-1}{2}]} \frac{(2j-1)!!}{(2j)!!} \sin^{2j}\theta_i\right),~m~\text{is odd}.
\end{cases} \label{eq:inres2}
\end{align}
To be explicit, the first few $I_n$'s are
\begin{subequations}
\label{eq:I0-I3}
\begin{align}
I_0(\theta_i)&=\frac{\pi }{2}-\theta_i,\\
I_1(\theta_i)&=\cos \left(\theta _i\right),\\
I_2(\theta_i)&=\frac{1}{4} \left(\pi-2 \theta _i+\sin \left(2 \theta _i\right) \right),\\
I_3(\theta_i)&=\frac{1}{12} \left(9 \cos \left(\theta _i\right)-\cos \left(3 \theta _i\right)\right).
\end{align}
\end{subequations}

When $\theta_s=\theta_d=0$ as in the infinite distance case, this can be further simplified to
\be \label{eq:i00res}
I_m(0,0)=
\frac{(m-1)!!}{m!!}\times\begin{cases}
\displaystyle  \pi,~~m\text{ is even},\\
\displaystyle  2,~~m\text{ is odd}.
\end{cases}
\ee
For the finite distance case however, $b/r_{s,d}~(i=s,d)$ is only small but nonzero. In this case, we can make an expansion of $I_n$ for small $b/r_{s,d}$ too by first using Eq. \eqref{eq:thetasddef} to expand $\theta_i$ in this limit. To the first few orders, we have
\begin{equation}
\label{eq:thetaexpansion}
\theta_i=\frac{b}{r_i}-\frac{b}{r_i^2}\left(\frac{c_1}{2}-\frac{a_1}{2 v^2}\right)+\frac{b^3}{6r_i^3}+\frac{b}{4r_i^3}\left(\frac{3 c_1^2-4 c_2}{2}-\frac{2a_1^2-2a_2+a_1c_1}{v^2}+\frac{2a_1^2}{2v^4}\right)+\mathcal{O}(\epsilon)^4
\end{equation}
where $\epsilon$ stands for either the infinitesimal $b/r_i$ or $M/b$.
Substituting this into the first few $I_n$ in Eq. \eqref{eq:I0-I3}, their expansions becomes \begin{subequations}
\label{eq:I0-I3expansion}
\begin{align}
I_0&=\frac{\pi }{2}-\frac{b}{r_i}+\frac{b}{r_i^2}\left(\frac{c_1}{2}-\frac{a_1}{2 v^2}\right)-\frac{b^3}{6r_i^3}-\frac{b}{4r_i^3}\left(\frac{3 c_1^2-4 c_2}{2}-\frac{2a_1^2-2a_2+a_1c_1}{v^2}+\frac{2a_1^2}{2v^4}\right)+\mathcal{O}(\epsilon)^4,\\
I_1&=1-\frac{b^2}{2 r_i^2}-\frac{b^2 \left(a_1-c_1 v^2\right)}{2 v^2 r_i^3}+\mathcal{O}(\epsilon)^3,\\
I_2&=\frac{\pi }{4}-\frac{b^3}{3 r_i^3}+\mathcal{O}(\epsilon)^4,\\
I_3&=\frac{2}{3}+\mathcal{O}(\epsilon)^4.
\end{align}
\end{subequations}

\end{document}